\documentclass[11pt]{article}
\newcommand{\be}{\begin{equation}}
\newcommand{\ee}{\end{equation}}
\newcommand{\bea}{\begin{eqnarray}}
\newcommand{\eea}{\end{eqnarray}}

\newcommand{\sech}{{\rm sech}}

\begin{document}
\vspace{0.5in}
\begin{center}
{\LARGE{Family of Potentials with Power-Law Kink Tails}} 
\end{center}

\begin{center}
{\LARGE{\bf Avinash Khare}} \\
{Physics Department, Savitribai Phule Pune University \\
 Pune 411007, India}
\end{center}

\begin{center}
{\LARGE{\bf Avadh Saxena}} \\ 
{Theoretical Division and Center for Nonlinear Studies, 
Los Alamos National Laboratory, Los Alamos, New Mexico 87545, USA}
\end{center}

\vspace{0.9in}
{\bf {Abstract:}}

We provide examples of a large class of one dimensional higher order field 
theories with kink solutions which asymptotically have a power-law tail 
either at one end or at both the ends. In particular, we  provide examples 
of a family of potentials which admit a kink as well as a mirror kink solution 
where all four ends of the two kinks have a power law tail or only the two 
extreme ends of the two kinks have a power law tail (while the two ends facing 
each other have an exponential tail) or the ends facing each other have a 
power law tail while the two extreme ends of the two kinks have an exponential 
tail. Further, we show that for a kink with a power law tail at either one end or 
at both the ends, there is no gap between the zero mode and the continuum 
of the corresponding stability equation. This is in contrast to the kinks with 
exponential tail at both the ends in which case there is always a gap between 
the zero mode and the continuum. 

\section{Introduction} 



Recently it was found that certain higher order field theories admit kink solutions
with a power law tail at either both the ends or a power law tail at one end and an 
exponential tail at the other end \cite{KCS, Chapter}. An example of the latter is the octic 
potential studied in the context of massless mesons \cite{Lohe}. This is in contrast 
to almost all the kink solutions that have been discussed in the last four decades 
where the kink solutions have an exponential tail at both the ends \cite{manton1, manton2}, 
the prototype being the celebrated  $\phi^4$ kink. The discovery of these power law 
kinks \cite{Gomes, Bazeia, Guerrero, Mello} has raised several interesting 
questions such as the strength and the range of the kink-kink (KK) and 
kink-antikink (K-AK) force \cite{gani1, manton3}, the possibility of 
resonances \cite{christov} and scattering 
\cite{gani2}, stability analysis of such kinks \cite{Gomes}, etc. From this 
perspective, it is worth noting that the celebrated Manton's method 
\cite{manton1, manton2} provides the answer for both the strength and the 
range of the KK and K-AK interactions in case 
they have an exponential tail at both the ends.

The study of higher order field theories, their attendant kink excitations as 
well as the associated kink interactions and scattering are important in a 
variety of physical contexts ranging from successive phase transitions 
\cite{KCS, Chapter, Gufan1,Gufan2} to isostructural phase transitions 
\cite{Pavlov} to models involving long-range interaction 
between massless mesons \cite{Lohe}, as well as from protein crystallization 
\cite{Boulbitch} to successive phase transitions presumably driving the 
late time expansion of the Universe \cite{Greenwood}. Thus, understanding 
kink behavior in these models provides useful insight 
into the properties of domain walls in materials, condensed matter, high energy 
physics, biology and cosmology, which also serves as one of our main motivations here. 

Recently several attempts have been made \cite{gani1, manton3, christov} to 
understand the nature of the KK and K-AK interaction in a $\phi^8$ model 
\cite{Lohe} which admits a kink and a mirror kink (and corresponding 
anti-kinks) which have exponential tails at the two extreme ends and power law
tails at the two adjoining ends. Very recently, following Manton's suggestion 
\cite{manton3}, KK and K-AK long range force has been calculated analytically 
for a one parameter family of potentials and then
compared with detailed numerical simulations \cite{cddgkks}. 

These studies raise several issues. Thus for example, what is the force between
a kink and the corresponding mirror kink in case both fall off with a power 
law tail at the two extreme ends while the tails of the two adjoining kinks 
have exponential tails? Secondly what is the force between a kink and the
corresponding mirror kink in case there is a power law tail at all the four ends? 
Besides, is there a conceptual difference in the stability analysis when there is 
an exponential tail at both the ends compared to when there is a power law tail 
at either both the ends or at one end? 

To facilitate such studies, in this paper we obtain kink solutions in a one
parameter family of potentials for which there is a power law tail at both the 
ends. The kink stability analysis for all these cases shows that the 
zero mode and the continuum start at the same energy (i.e., zero), that is, 
there is no gap between the zero mode and the continuum. This is  
in contrast to the kinks with exponential tails at both the ends where there 
is always a gap between the zero mode and where the continuum begins. 
As an illustration, in Appendix A we consider a one-parameter 
family of potentials which admit a kink solution with an exponential tail at 
both the ends. The stability analysis for these kink solutions reveals the
existence of a gap between the zero mode and the beginning of the continuum.
Besides, we construct a one-parameter family of kink and corresponding mirror 
kink solutions in models such that (i) the kink (and hence the mirror kink) has
a power law tail at both the ends (ii) the kink (and hence the mirror kink) 
has a power law tail at the two extreme ends and an exponential tail at the two
adjacent ends (iii) the two kinks have an exponential tail at the two extreme
ends while they have a power law tail at the two adjacent ends.
Unfortunately, in all these cases we are only able to obtain 
the kink solution in an implicit form. However we are able to obtain the
relevant asymptotic tails in all these cases. 

The plan of the paper is as follows. In Sec. II we set up the  notations and 
illustrate as to when one can have a kink solution with power law tail and when 
one can have a kink solution with exponential tail. Besides, we consider the
case of two adjoining kinks and point out the various possible forms for the
kink tails in the two adjoining kink solutions. 
In Sec. III we present a one parameter
family of potentials of the form $(a^2 -\phi^2)^{2n+2}$, $n = 1, 2, 3...$, all
of which admit a kink solution from $-a$ to $+a$ with a power law tail at both
the ends. Unfortunately, all these kink solutions are expressed in an implicit 
form. The stability analysis of these kink solutions is performed in
Sec. 3.5 where we show that there is no gap between the zero mode and the
continuum. In rest of the paper we
construct one parameter family of potentials corresponding to the various 
possible forms for the two adjoining kinks. 
In Sec. IV we present a one-parameter family of potentials of
the form $\phi^{2n+2} (a^2 -\phi^2)^2$, $n = 1, 2, 3,...$, all of which admit
a kink solution from $0$ to $a$ and a mirror kink from $-a$ to $0$ (and the
corresponding anti-kinks). For all these kink solutions, while around 
$\phi = 0$ one has a power law tail, around $\phi = a$ one has only 
an exponential tail.
In Sec. V we present a one-parameter family of potentials of the
form $\phi^{2}(a^2 -\phi^2)^{2n+2}$,~ $n = 1,
2, 3,...$, which admit
a kink solution from $0$ to $a$ and a mirror kink from $-a$ to $0$ (and the
corresponding anti-kinks) and for which while round $\phi = 0$ one has an 
exponential tail, around $\phi = a$ one has a power law tail. Finally in 
Sec. VI we present a one-parameter family of potentials of the
form $\phi^{2n+2}(a^2 -\phi^2)^{4}$,~ $n = 2, 3,...$, which admit
a kink solution from $0$ to $a$ and a mirror kink from $-a$ to $0$ (and the
corresponding anti-kinks). For these kinks, there is power law tail 
around both $\phi = 0$ as well as $\phi = a$ such that the power  law tail 
around $\phi = 0$ has a slower fall off (i.e. it goes like $x^{-1/n}$ as 
$x \rightarrow \pm \infty$ where $n = 2,3,4,...$) compared to the power law tail
around $\phi = a$ (which always goes like $x^{-1}$ as 
$x \rightarrow \infty$). For completeness, we also consider
in Sec. VI a one-parameter family of potentials of the
form $\phi^{4}(a^2 -\phi^2)^{2n+2}$ and $n = 2, 3,...$, which admit
a kink solution from $0$ to $a$ and a mirror kink from $-a$ to $0$ (and the
corresponding anti-kinks) and for which around $\phi = 0$ as well as around
$\phi = a$ one has a power law tail such that the power law tail around 
$\phi = a$ has a slower fall off (i.e. it goes like $x^{-1/n}$ as 
$x \rightarrow \infty$ where $n = 2, 3, 4,...$) compared to the power law tail
around $\phi = 0$ (which always goes like $x^{-1}$ as $x \rightarrow -\infty$). 
In Sec. VII we summarize our main results and point out some of the open 
problems. 

In Appendix A we present a one parameter family 
of potentials of the form $\phi^2 (a^{2n}-\phi^{2n})^2$ with 
$n = 1, 2, 3,...$, which 
admits a kink solution from $0$ to $a$ and a mirror kink from $-a$ to 
$0$ both of which have an exponential tail at both the ends. For the entire 
family, we obtain the kink solutions in an explicit form, perform the 
linear stability analysis and show 
that there is always a gap between the zero mode and the continuum.
In Appendix B, for completeness we present a one parameter
family of potentials of the form $|a^2 -\phi^2|^{2n+1}$, $n = 1, 2, 3...$, all
of which admit a kink solution from $-a$ to $+a$ with a power law tail at both
the ends. In the special case of $n = 1$ we obtain the kink solution 
analytically. Using this kink solution we then obtain the zero mode and also
carry out the linear stability analysis and show explicitly that there is no gap
between the zero mode and the continuum.
In Appendices C to E we provide details of the calculation performed in 
Sec. IV, Sec. V and Sec. VI, respectively. For completeness, in Appendix F we 
present a one-parameter family of potentials of the
form $(a^2-\phi^2)^{2} (b^2-\phi^2)^{2n+2}$ which admit three kink solutions. 
Two of these, i.e. from $a$ to $b$ (and a mirror kink from $-b$ to $-a$)
have a power law tail at one end and an exponential tail at the other end 
while the kink from $-a$ to $a$ has an exponential tail at both the ends. Finally, 
in Appendix G we present a one parameter family of potentials of the
form $(a^2-\phi^2)^{2n+2} (b^2-\phi^2)^{2} $ which admit three kink solutions. 
Two of these, i.e. from $a$ to $b$ (and a mirror kink from $-b$ to $-a$)
have an exponential tail at one end and a power law tail at the other end 
while the kink from $-a$ to $a$ has a power law  tail at both the ends. 
 
\section{Formalism}

Consider a relativistic neutral scalar field theory in $1+1$ dimensions with
the Lagrangian density
\be\label{2.1}
{\mathcal L} = \frac{1}{2}\left(\frac{\partial \phi}{\partial t}\right)^2  
- \frac{1}{2}\left(\frac{\partial \phi}{\partial x}\right)^2 - V(\phi)\,,
\ee
which leads to the equation of motion
\be\label{2.2}
\left(\frac{\partial^2 \phi}{\partial^2 t}\right)^2  
- \left(\frac{\partial^2 \phi}{\partial^2 x}\right)^2 = - \frac{dV}{d\phi}\,.
\ee
We assume that the potential $V(\phi)$ is smooth and non-negative. Thus 
$V(\phi)$ attains its global minimum value of $V = 0$ for one or more values of 
$\phi$ which are the global minima of the theory. We shall choose $V(\phi)$ 
such that it has two or more global minima so that one has static kink and 
anti-kink solutions interpolating between the two adjoining global minima 
as $x$ increases
from $-\infty$ to $+\infty$. While the field equation for a static kink is a 
second order ODE, it can be reduced to a first-order ODE using the so 
called Bogomolny technique. The first order ODE is given by
\be\label{2.3}
\frac{d\phi}{dx} = \pm \sqrt{2V(\phi)}\,.
\ee
The corresponding static kink energy (which also equals the corresponding 
anti-kink energy) and which is also referred to as kink mass is given by
\be\label{2.4}
M_k = \int_{-\infty}^{\infty} \bigg (\frac{1}{2} \left(\frac{d\phi}{dx}\right)^2  
+V(\phi) \bigg )\, dx\,.
\ee
In view of the first order Eq. (\ref{2.3}), the kink mass $M_k$ takes a 
simpler form
\be\label{2.5}
M_k = \int_{\phi_a}^{\phi_b} \sqrt{2V(\phi)}\, d\phi\,,
\ee
where as $x$ goes from  -$\infty$ to +$\infty$, the kink solution goes from
one minimum $\phi_a$ to the adjacent minimum $\phi_b$. 

One can perform the linear stability analysis by considering
\be\label{2.6}
\phi(x,t) = \phi_k(x) + \eta(x) e^{i\omega t}\,,
\ee
where $\phi_k$ is the kink solution of the first order Bogomolny 
Eq. (\ref{2.3}). On substituting $\phi(x,t)$ as given by Eq. (\ref{2.6}) in the
field Eq. (\ref{2.2}) and retaining terms of order $\eta$, it is easily shown 
that $\eta(x)$ satisfies a Schr\"odinger-like equation
\be\label{2.7} 
-\frac{d^2 \eta}{dx^2} + \frac{d^2 V(\phi)}{d\phi^2} \Big |_{\phi = \phi_{k}(x)} 
\eta(x) = \omega^2 \eta\,.
\ee
Here $\phi_{k}(x)$ denotes the corresponding kink (or anti-kink) solution. It
is well known that the stability Eq. (\ref{2.7}) always admits a zero-mode 
solution, i.e.
\be\label{2.8}
\omega_0 = 0\,,~~\eta_0 (x) = \frac{d\phi_k(x)}{dx}\,,
\ee
where $\eta_0(x)$ is nodeless, thereby guaranteeing the linear stability of the 
kink solution of any theory. 

In this paper we shall choose various forms of the potential $V(\phi)$ and 
obtain the corresponding kink and antikink solutions. While majority of
the known kink solutions have an exponential tail, only few kink solutions have
so far been constructed having a power law tail. The recipe for 
constructing kink solutions with a power law tail or an exponential tail is clear
and well known. Since a kink solution has finite energy it implies that the
solution must approach one of the minima (vacua) $\phi_0$ of the theory as
$x$ approaches either +$\infty$ or -$\infty$. If the lowest non-vanishing 
derivative of the potential at the minimum has order $m$, then by Taylor
expanding the potential at the minimum and writing the field close to it as
$\phi = \phi_0 + \eta$, one finds that the self-dual first order equation
in $\eta$ implies that (assuming that the potential vanishes at the minimum)
\be\label{2.9}
\frac{d\eta}{dx} \propto \eta^{m/2}\,.
\ee
Thus if $m =2$ then $\eta \propto e^{-\alpha x}$ while if $m > 2$ then 
$\eta \propto 1/x^{2/(m-2)}$. We shall use this recipe to construct 
several one parameter family of potentials with various possible forms of
power law and exponential tails.    

Using the above recipe one can construct models which can give us kink 
solutions with either a power law tail or an exponential tail. Even more importantly,
in most of the cases discussed in this paper, we construct models which give 
rise to a kink and a mirror kink with 
various possible options for the kink tails. Let us denote the two
adjoining kink solutions as kink 1 and kink 2 and without loss of generality
we will assume that kink 1 is to the left of kink 2. One has two kink 
tails corresponding to the kink solution one (which we denote by $K_{1L}$
and $K_{1R}$) and two kink tails $K_{2L}$ and $K_{2R}$ for solution two. In 
Table 1 we give all 8 possible forms of the kink tails. The previously well studied 
case is when all the four tails (i.e. two tails of the first kink and the two 
tails of the second kink) have exponential fall off which we denote by  
$K_{1Le}, K_{1Re}, K_{2Le}, K_{2Re}$ and for simplicity we will denote 
such tails simply as $eeee$. On the other hand, the recent study by several 
groups \cite{gani1, manton3, christov, cddgkks} have concentrated on 
the case when the kink tails have the form $K_{1Le}, K_{1Rp}, K_{2Lp}, 
K_{2Re}$ and  we shall denote this possibility as $eppe$. We remark that
for both of these cases, a kink and the corresponding mirror kink are possible. 
In fact, there are two other possibilities for which kinks and corresponding 
mirror kinks are also possible and these are of the form $peep$ and $pppp$. 
Correspondingly, for the other four possibilities shown in Table I, one necessarily 
has to consider non-mirror kinks since for them the kink tails are of the form 
$eeep$ (and $peee$, and without loss of generality one can consider one of 
these two possibilities) and $pppe$ (and $eppp$, and again without loss of 
generality we can consider only one of these two possibilities). 

\begin{table}[h!]
\centering
\caption{Eight different cases of kink tail configurations. Here e denotes an 
exponential tail and p denotes a power law tail (see text for details).}
\vskip 0.3 truecm 
\begin{tabular}{ cccc }
 \hline 
 K$_{1L}$ & K$_{1R}$ & K$_{2L}$ & K$_{2R}$ \\
 \hline
  e & e & e & e \\
  e & p & p & e \\
  p & e & e & p \\
  p & p & p & p \\
  e & e & e & p \\ 
  p & e & e & e \\ 
  p & p & p & e \\ 
  e & p & p & p \\
\hline 
\end{tabular}
\end{table} 

We now consider the various possible forms of the kink tails mentioned above 
and construct a one-parameter family of potentials corresponding to each of 
these six possibilities. To initiate the discussion, in the next section we 
consider models admitting only one kink solution with a power law tail at both 
the ends. 

\section{Power-law tails at both ends} 

In this section we present a one-parameter family of potentials 
\be\label{3.1} 
V(\phi) = \lambda^2(\phi^2-a^2)^{2n+2} \,, ~~~ n=1,2,3, ... \,. 
\ee 
which admits a kink solution with a power law tail at both the ends. 
This potential has degenerate minima at $\phi = \pm a$ and $V(\phi = \pm a)
= 0$. We therefore look for a kink solution which oscillates from -$a$ to +$a$
as $x$ goes from -$\infty$ to +$\infty$. Thus for the kink solutions discussed
in this section, $\phi^2 \le a^2$. 
In these cases explicit analytic solutions are, unfortunately not possible 
and we can only find implicit kink solutions. From the latter we can obtain 
how a kink profile falls off as $x\rightarrow \pm\infty$.  We will first 
discuss explicitly the case $n=1$ (i.e. the $\phi^8$ field theory) and $n=2$ 
(i.e. the $\phi^{12}$ field theory) which have already been discussed in 
\cite{KCS} and then generalize to arbitrary $n$. 

\subsection{Case I: $n=1$}

For $n =1$, Eq. (\ref{3.1}) leads to the potential
\be\label{3.2} 
V(\phi) = \lambda^2 (\phi^2 - a^2)^4 \,. 
\ee 

On using Eq. (\ref{2.3}), the self-dual first order 
equation is 
\be\label{3.3} 
\frac{d\phi}{dx} = \pm \sqrt{2} \lambda (a^2 - \phi^2)^2\,. 
\ee 
This is easily integrated by using the identity
\be\label{3.4}
\int \frac{d\phi}{(a^2-\phi^2)^{n}} = \frac{(2n-3)}{2(n-1)a^2} 
\int \frac{d\phi}{(a^2-\phi^2)^{n-1}} + 
\frac{\phi}{2(n-1)(a^2-\phi^2)^{n-1} a^2}\,. 
\ee 
We find
\be\label{3.5} 
\mu x = \frac{2a\phi}{a^2-\phi^2} + \ln\left(\frac{a+\phi}{a-\phi}\right) \,, ~~~ \mu = 4\sqrt{2}\lambda a^3 \,. 
\ee 
From here it is straightforward to show that 
\be\label{3.6} 
\lim_{x \rightarrow -\infty} ~\phi(x) = - a + \frac{a}{-\mu x} + ... \,.  ~~~  
\lim_{x \rightarrow +\infty} ~\phi(x) = a - \frac{a}{\mu x} + ... \,. 
\ee 
Notice that the precise expressions for the kink tails around both $\phi = +a$ 
and $-a$ are entirely decided by the first term on the right hand side of 
Eq. (\ref{3.5}), i.e. by the term $\frac{2a\phi}{(a^2-\phi^2)}$. 

\subsection{Case II: $n=2$}
  
For $n =2$, Eq. (\ref{3.1}) leads to the potential
\be\label{3.7} 
V(\phi) = \lambda^2 (\phi^2 - a^2)^6 \,. 
\ee 
On using the fact that for the kink solution between $-a$ and +$a$, 
$\phi^2 < a^2$ and the identity (\ref{3.4}), the kink solution turns out to be  
\be\label{3.8} 
\mu x = \frac{4 a^3 \phi} {3(a^2-\phi^2)^2} + \frac{2\phi a}{(a^2 -\phi^2)} 
+ \ln\left(\frac{a+\phi}{a-\phi}\right) \,, 
~~~ \mu = \frac{16}{3}\sqrt{2}\lambda a^5 \,. 
\ee 
Asymptotically, we find 
\be\label{3.9} 
\lim_{x \rightarrow -\infty} ~\phi(x) = -a + \frac{a}{\sqrt{-3\mu x}} + ... \,, ~~~
\lim_{x \rightarrow +\infty} ~\phi(x) = a - \frac{a}{\sqrt{3\mu x}} + ... \,. 
\ee 
Again observe that the kink tails around both $\phi = +a$ and $-a$ are entirely
given by the first term on the right hand side of Eq. (\ref{3.8}), i.e.  
by the term $\frac{4a^3\phi}{3(a^2-\phi^2)^2}$. 

\subsection{General Case: Arbitrary $n$} 

The generalization to arbitrary $n$ is now straightforward.
In this case the potential is
\be\label{3.10} 
V(\phi) = \lambda^2(\phi^2-a^2)^{2n+2} \,. 
\ee
On using the identity (\ref{3.4}) repeatedly (also see the integral 2.149(3) 
of \cite{gr}]), one can show that 
\bea\label{3.11} 
&&\mu x = \frac{2^{n}(n-1)!}{(2n-1)!!} 
\bigg [\frac{a^{2n-1}\phi }{(a^2-\phi^2)^n} \nonumber \\
&& + \sum_{k=1}^{n-1} \frac{ (2n-1)(2n-3)...(2n-2k+1)}{2^{k}(n-1)(n-2)...(n-k)} 
\frac{a^{2n-2k-1}\phi}{(a^2-\phi^2)^{n-k}} \bigg ] 
+ \ln\left(\frac{a+\phi}{a-\phi}\right) \,, \nonumber \\ 
\eea 
where 
\be\label{3.12} 
\mu = \frac{2^{n+3/2}\lambda a^{2n+1} n!} {(2n-1)!!}  \,. 
\ee 
Asymptotically, the leading contribution as $x\rightarrow\pm\infty$ comes from 
the first term on the right hand side of Eq. (\ref{3.11}).  We find that 
\bea\label{3.13} 
&&\lim_{x \rightarrow -\infty} ~\phi(x) = -a 
+ \frac{a}{\left(-\frac{(2n-1)!! ~\mu x}
{(n-1)!}\right)^{1/n}} + ... \,, \nonumber \\
&&\lim_{x \rightarrow +\infty} ~\phi(x) = a 
- \frac{a}{\left(\frac{(2n-1)!! ~\mu x}{(n-1)!}\right)^{1/n}} + ... \,. 
\eea 
As expected, for $n=1,2$ these results for the kink tails agree with those 
given by Eqs. (\ref{3.6}) and (\ref{3.9}), respectively. Thus the 
kink tail around both $\phi = +a$ and $-a$ goes like $x^{-1/n}$ and hence 
as $n$ increases, the kink-antikink interaction becomes highly nonlinear.  

For completeness, in Appendix B we consider kink solutions for the 
one-parameter family of potentials 
\be\label{3.1a}
V(\phi) =\lambda^2 |a^2 -\phi^2|^{2n+1}\,,~~n = 1, 2, ...\,.
\ee
For the special case of $n = 1$ and $n = 3$, we analytically obtain  
the kink solution in an explicit form.

\subsection{Kink Mass}

Using Eq. (\ref{2.5}) one can immediately estimate the kink mass for the
entire family of potentials as given by Eq. (\ref{3.1}). We find that 
\be\label{3.14}
M_k(n) = \frac{2^{4n+3/2} \lambda a^{4n+1} [(2n)!]^2}{(4n+1)!}\,.
\ee
For example, while
\be\label{3.15}
M_k(n = 1) = \frac{16 \sqrt{2}}{15} \lambda a^{5}\,,~~
M_k(n = 2) = \frac{256 \sqrt{2}}{315} \lambda a^{9}\,.
\ee
We note that for $a = 1$, the kink mass decreases as $n$ increases.

\subsection{Linear Stability Analysis} 

Let us consider the linear stability of the kink solutions discussed in this 
section. For all these kink solutions we note that  in
the $\lim_{x \rightarrow \pm \infty}~ \phi(x) \rightarrow \pm a$, so that for 
all these kink solutions
\be\label{3.14a} 
\lim_{x \rightarrow \pm \infty}~  (a^2-\phi^2)  \rightarrow  0 \,. 
\ee 
Now the potential $V(x)$ in the stability  Eq. (\ref{2.7}) is given by
\be\label{3.15a} 
V(x) = d^2 V(\phi)/d\phi^2|_{\phi = \phi_k(x)} \,. 
\ee 
But since all the potentials $V(\phi)$ discussed in this section are of the form
\be\label{3.16} 
V(\phi) = (a^2-\phi^2)^{2n+2} \,, ~~~ n= 1,2,3... \,, 
\ee 
hence on differentiating this $V(\phi)$ twice with respect to $\phi$
it follows that at least a factor of $(a^2-\phi^2)^{2}$ will always remain.
But since the kink solution goes to $-a$ or +$a$ as $x =\rightarrow \mp \infty$
hence the potential $V(x)$ as given by Eq. (27) vanishes as 
$x \rightarrow \pm \infty$. 
Using the standard results of quantum mechanics in one dimension, it then
follows that for $\omega^2 \ge 0$, the spectrum of the Schr\"odinger-like 
Eq. (7) for the above class of potentials is continuous, with 
$\omega^2 = 0$ being the beginning of the continuum. But for
any kink solution one knows that there exists a nodeless zero mode at 
$\omega^2 = 0$ which guarantees the linear stability of the kink solution. 
Thus there is {\it no gap} between the zero mode and the continuum in the case of 
the kink solutions discussed in this section, all of which have a power law 
tail. 

This argument is rather general and goes through for 
any kink solution for which there is a power law tail at either both the ends 
or at one of the ends because in all these cases the continuum always begins at 
$\omega^2 = 0$. We have checked this for all the kink solutions with a 
power law tail discussed 
in this paper and in all these cases we find that there is no gap 
between the zero mode and the continuum. In contrast, for kinks with an 
exponential tail at both the ends, there is {\it always a gap} between the zero mode 
and the continuum. As an illustration, in Appendix A 
we demonstrate it for the kink solutions of the one-parameter family of
potentials of the form $\phi^2(\phi^{2n}-a^{2n})^2$ where $n = 1, 2, 3,...$.

\section{Potentials admitting kink and mirror kink solutions with tails of 
the form $e~p~p~e$}

In the previous section we considered kink solutions of the form $pppp$. 
In this section we present a one parameter family of potentials of the form
\be\label{4.1}
V(\phi) = \lambda^2 \phi^{2n+2}( a^2-\phi^2)^2  \,, ~~~ n = 1, 2, 3, ... ~.
\ee
This potential has degenerate minima at $\phi = 0, \pm a$ and 
$V(\phi = 0, \pm a) = 0$ which admits 
a kink from $0$ to $a$ and a mirror kink from $-a$ to $0$ and both the kinks
have a power law tail at one end and an exponential tail at the other end.
We  look for a kink solution which oscillates from $0$ to +$a$
as $x$ goes from $-\infty$ to +$\infty$. Thus for the kink solution 
$\phi^2 \le a^2$. 
We consider the two cases of odd $n$ (i.e. potentials of the form $\phi^{4n}$) 
and even $n$ (i.e. potentials of the form $\phi^{4n+2}$) separately. 
Unfortunately, in all these cases, explicit analytic solutions are not 
possible and we can only find implicit kink solutions. From the latter we 
can obtain how a kink profile falls off as $x\rightarrow\pm\infty$.

\subsection{$\phi^{4n}$ potentials} 

\noindent Consider the class of potentials 
\be\label{4.2} 
V(\phi) = \lambda^2 \phi^{4n}( a^2-\phi^2)^2  \,, ~~~ n = 1, 2, 3, ... ~.
\ee
We will first discuss explicitly the case $n=1$ (i.e. the $\phi^8$ field theory) 
and $n=2$ (i.e. the $\phi^{12}$ field theory) which have already been discussed in
\cite{KCS} and then generalize to arbitrary $n$. 

\subsubsection{Case I: $n=1$}   

On using Eq. (\ref{2.3}), the self-dual first order equation is 
\be\label{4.3a} 
\frac{d\phi}{dx} = \pm \sqrt{2} \lambda \phi^2 (a^2 - \phi^2)\,. 
\ee 
In this case
\be\label{4.3} 
\sqrt{2} \lambda x = \int \frac{d\phi}{\phi^2(a^2-\phi^2)} \,.
\ee 
The integrand on the right hand side can be written as
\be\label{4.4}
\frac{1}{a^2 \phi^2} + \frac{1}{a^2 (a^2 - \phi^2)}\,.
\ee

This is easily integrated with the solution 
\be\label{4.5} 
\mu x = -\frac{2a}{\phi} + \ln\frac{a+\phi}{a-\phi} \,, 
~~~ \mu=2\sqrt{2} \lambda a^3 \,. 
\ee 
Asymptotically, 
\be\label{4.6} 
\lim_{x \rightarrow -\infty} ~\phi(x) = \frac{2a}{-\mu x} \,, 
~~~ \lim_{x \rightarrow \infty} ~\phi(x) = a - 2a e^{-\mu x-2} \,. 
\ee 
Thus the kink tail around $\phi = 0$ is entirely determined by the 
first term on the right hand side of Eq. (\ref{4.5}), i.e. the term 
$-\frac{2a}{\phi}$.

\subsubsection{Case II: $n=2$} 

On using Eq. (\ref{2.3}), the self-dual first order equation is 
\be\label{4.7} 
\sqrt{2} \lambda x = \int \frac{d\phi}{\phi^4(a^2-\phi^2)} \,. 
\ee 
The integrand on the right hand side can be evaluated using partial fractions
and it can be shown to be
\be\label{4.8}
\frac{1}{a^2 \phi^4} + \frac{1}{a^4 \phi^2} + \frac{1}{a^4 (a^2 - \phi^2)}\,.
\ee
This is easily integrated with the solution 
\be\label{4.9}
\mu x = - \frac{2a^3}{3\phi^3} -\frac{2a}{\phi} + \ln\frac{a+\phi}{a-\phi} \,, 
~~~ \mu=2\sqrt{2} \lambda a^5 \,. 
\ee 
Asymptotically, 
\be\label{4.10} 
\lim_{x \rightarrow-\infty} ~\phi(x) = \frac{2^{1/3}a}{(-3\mu x)^{1/3}} \,,
 ~~~ \lim_{x \rightarrow \infty} ~\phi(x) = a - 2a e^{-\mu x-2-2/3} \,. 
\ee 
Thus the kink tail around $\phi = 0$ is again entirely determined by the 
first term on the right hand side of Eq. (\ref{4.9}), i.e. by the term
-$\frac{2 a^3}{3\phi^3}$. Further, on comparing the kink tail around $\phi 
= a$ (i.e. as $x \rightarrow +\infty$) in the case of $n = 1$ and $n = 2$ 
as given by Eqs. (\ref{4.6}) and (\ref{4.10}) respectively, we find that in 
both the cases, the kink tail has an exponential form 
\be\label{4.11}
\lim_{x \rightarrow \infty}~ \phi(x) = a - 2a e^{-\mu x - K}\,,
\ee
where the constant $K$ equals $2$ in the $n = 1$ case while it is $2+2/3=8/3$ in
the $n = 2$ case, whereas the rest of the behaviour is same in both the cases.

\subsubsection{Case III: General $n$} 

The generalization to arbitrary $n$ is now straightforward.
On using Eq. (\ref{2.3}), the self-dual first order equation is 
\be\label{4.12}
\sqrt{2} \lambda x = \int \frac{d\phi}{\phi^{2n}(a^2-\phi^2)} \,. 
\ee 
The integrand on the right hand side can be evaluated using partial fractions
and it can be shown to be
\be\label{4.13}
\sum_{k=1}^{n} \frac{1}{a^{2k} \phi^{2n+2-2k}} 
+\frac{1}{a^{2n} (a^2 - \phi^2)}\,.
\ee
This is easily integrated with the solution 
\be\label{4.14} 
\mu x = - \frac{2a^{2n-1}}{(2n-1)\phi^{2n-1}} 
-\sum_{k =2}^{n} \frac{2a^{2n-2k+1}}{(2n-2k+1)\phi^{2n-2k+1}} 
+\ln\frac{a+\phi}{a-\phi} \,,
\ee
where
\be\label{4.15}
\mu=2\sqrt{2} \lambda a^{2n+1} \,. 
\ee 
Asymptotically, 
\bea \label{4.16}
&&\lim_{x \rightarrow -\infty} ~\phi(x) = \frac{2^{1/{(2n-1)}}a}
{[-(2n-1)\mu x]^{1/(2n-1)}} \,, \nonumber \\ 
&&\lim_{x \rightarrow \infty} ~\phi(x) = a - 2a e^{-\mu x-B_n} \,,  
\eea 
where $B_n = 2 +2/3 + 2/5 + ... + 2/(2n-1)$. 

As expected, for $n=1,2$ these results for the kink tail agree with those 
given by Eqs. (\ref{4.6}) and (\ref{4.10}), respectively. Thus for the class of 
potentials as given by Eq. (\ref{4.2}), the 
kink tail around $\phi = +a$ goes like $e^{-\mu x}$ while it goes like 
$x^{-1/(2n-1)}$ around $\phi = 0$ and hence 
as $n$ increases, the kink-antikink interaction becomes highly nonlinear.  

\subsection{$\phi^{4n+2}$ potentials} 

Let us now consider the class of potentials 
\be\label{4.17} 
V(\phi) = \lambda^2 \phi^{4n+2}( a^2-\phi^2)^2  \,, ~~~ n = 1, 2, 3, ... ~.
\ee
First we will consider the $n=1$ case (which is already discussed in 
\cite{KCS}) and then mention the results for the kink tail for the general 
$n$ case. The details about the $n=2$ as well as general $n$ case are given in
Appendix C.

\subsubsection{Case I: $n=1$}   

On using Eq. (\ref{2.3}) the self-dual first order equation is
\be\label{4.18} 
\sqrt{2} \lambda x = \int \frac{d\phi}{\phi^3(a^2-\phi^2)} \,, 
\ee
The integrand on the right hand side can be evaluated using partial fractions
and it can be shown to be
\be\label{4.19}
\frac{1}{a^2 \phi^3} + \frac{1}{a^4 \phi} + \frac{\phi}{a^4 (a^2 - \phi^2)}\,.
\ee
This is easily integrated with the solution 
\be\label{4.20} 
\mu x = -\frac{a^2}{\phi^2} + \ln\frac{\phi^2}{a^2-\phi^2} \,, ~~~ \mu=2\sqrt{2} \lambda a^4 \,. 
\ee 
Asymptotically, 
\be\label{4.21} 
\lim_{x \rightarrow -\infty} ~\phi(x) = \frac{a}{\sqrt{-\mu x}} \,, 
~~~ \lim_{x \rightarrow +\infty} ~\phi(x) = a - \frac{a}{2} e^{-\mu x-1} \,. 
\ee 
Thus the kink tail around $\phi = 0$ is entirely determined by the 
first term on the right hand side of Eq. (\ref{4.20}), i.e. by the term
-$\frac{a^2}{\phi^2}$. 

Similar to the $n = 1$ case, one can obtain results for the $n = 2$ case
and then generalize to arbitrary $n$, which are discussed in Appendix C. As
shown in Appendix C for arbitrary $n$, while
the kink tail falls off like $e^{-\mu x}$ around $\phi = a$, it falls off like 
$x^{-1/2n}$ around $\phi = 0$. 

On combining the results of Sec. 4.1 and 4.2, we then conclude that for the
one-parameter family of potentials as given by Eq. (\ref{4.1}) for arbitrary
$n$, while the kink tail falls off like $e^{-\mu x}$ around $\phi = a$, it 
falls off like $x^{-1/n}$ around $\phi = 0$. 

\subsection{Kink Mass}

Using Eq. (\ref{2.5}) one can immediately estimate the kink mass for the
entire family of potentials as given by Eq. (\ref{4.1}). We find that 
\be\label{4.31}
M_k = \frac{2^{3/2} \lambda a^{n+4}}{(n+2)(n+4)}\,,~~~~n = 1, 2, 3, ...\,.
\ee
For example, while
\be\label{4.32}
M_k(n = 1) = \frac{2\sqrt{2}}{15} \lambda a^{5}\,,~~
M_k(n = 2) = \frac{2\sqrt{2}}{24} \lambda a^{6}\,.
\ee
We note that for $a = 1$, the kink mass decreases as $n$ increases. 

\section{Potentials admitting kink and mirror kink solutions with tails of 
the form $p~e~e~p$}

In this section we present a one parameter family of potentials of the form
\be\label{5.1}
V(\phi) = \lambda^2 \phi^{2}( a^2-\phi^2)^{2n+2}  \,, ~~~ n = 1, 2, 3, ... ~.
\ee
These potentials have degenerate minima at $\phi = 0\,, \pm a$ and 
$V(\phi = 0, \pm a) = 0$ and admit 
a kink from $0$ to $a$ and a mirror kink from $-a$ to $0$ and both the kinks
have a power law tail at one end and an exponential tail at the other end.
We  look for a kink solution which oscillates from $0$ to +$a$
as $x$ goes from $-\infty$ to +$\infty$. Thus for the kink solution 
$\phi^2 \le a^2$. In these cases explicit analytic solutions are not possible 
and we can only 
find implicit kink solutions. From the latter we can obtain 
how a kink profile falls off as $x\rightarrow\pm\infty$.  

We will first discuss  the case $n=1$ (i.e. the $\phi^{10}$ field theory) 
which has already been discussed in \cite{KCS} and then mention results about
the kink tail for arbitrary $n$. The details about $n=2$ (i.e. the $\phi^{14}$ 
field theory) and the generalization to arbitrary $n$ are given in Appendix D. 

\subsection{Case I: $n=1$}   

On using Eq. (\ref{2.3}), the self-dual first order equation is 
\be\label{5.2} 
\frac{d\phi}{dx} = \pm \sqrt{2} \lambda \phi (a^2 - \phi^2)^2\,. 
\ee 
In this case
\be\label{5.3} 
\sqrt{2} \lambda x = \int \frac{d\phi}{\phi (a^2-\phi^2)^2} \,.
\ee 
The integrand on the right hand side can be written as partial fractions 
\be\label{5.4}
\frac{\phi}{a^2 (a^2 - \phi^2)^2} + \frac{\phi}{a^4 (a^2 - \phi^2)}
+\frac{1}{a^4 \phi}\,.
\ee
This is easily integrated with the solution 
\be\label{5.5} 
\mu x = \frac{a^2}{a^2 - \phi^2} +\ln\frac{\phi^2}{a^2-\phi^2}\,, 
~~~ \mu=2\sqrt{2} \lambda a^4 \,. 
\ee 
Asymptotically, 
\be\label{5.6} 
\lim_{x \rightarrow -\infty} ~\phi(x) = a e^{\mu x /2-1} \,, 
~~~ \lim_{x \rightarrow\infty} ~\phi(x) = a - \frac{a}{2\mu x} \,. 
\ee 
Thus the kink tail around $\phi = a$ is entirely determined by the 
first term on the right hand side of Eq. (\ref{5.5}), i.e, the term 
$\frac{a^2}{(a^2 -\phi^2)}$.

Similar to the $n = 1$ case, one can obtain results for the $n = 2$ case
and then generalize to arbitrary $n$, which are discussed in Appendix D. As
shown in Appendix D for arbitrary $n$, while the kink tail falls off like $e^{-\mu 
x /2}$ around $\phi = 0$, it falls off like $x^{-1/n}$ around $\phi = a$.

\subsection{Kink Mass}

Using  Eq. (\ref{2.5}) one can immediately estimate the kink mass for the
entire family of potentials as given by Eq. (\ref{5.1}). We find that 
\be\label{5.16}
M_k = \frac{\lambda a^{2n+4}}{\sqrt{2}(n+2)}\,,~~~~n = 1, 2, 3,,,\,.
\ee
For example, while
\be\label{5.17}
M_k(n = 1) = \frac{1}{3\sqrt{2}} \lambda a^{6}\,,~~
M_k(n = 2) = \frac{1}{4\sqrt{2}} \lambda a^{8}\,.
\ee
Thus for $a = 1$, the kink mass decreases as $n$ increases.

\section{Potentials admitting kink and mirror kink solutions with tails of 
the form $p~p~p~p$}

In this section we discuss a family of potentials of the form
\be\label{6.1}
V(\phi) = \lambda^2 \phi^{2m+2}( a^2-\phi^2)^{2n+2}  \,, 
~~~ n, m = 1, 2, 3, ... ~\,.
\ee
These potentials have degenerate minima at $\phi = 0, \pm a$ and 
$V(\phi = 0, \pm a) = 0$ and admit 
a kink from $0$ to $a$ and a mirror kink from $-a$ to $0$ and both the kinks
have a power law tail at both the ends.
We  look for a kink solution which oscillates from $0$ to +$a$
as $x$ goes from $-\infty$ to +$\infty$, respectively. Thus for the kink 
solution $\phi^2 \le a^2$. In these cases explicit analytic solutions are 
not possible and we can only 
find implicit kink solutions. From the latter we can obtain 
how a kink profile falls off as $x\rightarrow\pm\infty$.  

In these models depending on if $m < n$ ($m > n$), one can have kink
solutions for which the power law tail around $\phi = a$ has faster 
(slower) asymptotic fall off compared to the power law tail around $\phi = 0$. 
We shall now consider a one-parameter family of potentials of both types 
(i.e. where $m < n$ and $m > n$).

\subsection{Models Where Kink Tail Around $\phi = \pm a$ has a Slower Asymptotic 
Fall off Compared to the Tail Around $\phi = 0$}

For simplicity, let us consider the potentials 
\be\label{6.2}
V(\phi) = \lambda^2 \phi^{4}( a^2-\phi^2)^{2n+2}\,,~~~~n = 2,3 ...~\,.
\ee
Similar arguments also hold good in case $\phi^{4}$ in Eq. (\ref{6.2}) is 
replaced by the potential $\phi^{2m+2}$ with $m < n$.
This potential has degenerate minima at $\phi = 0, \pm a$ and 
$V(\phi = 0, \pm a) = 0$ which admit 
a kink from $0$ to $a$ and a mirror kink from $-a$ to $0$ and both kinks
have a power law tail at both the ends.
We  look for a kink solution which oscillates from $0$ to +$a$
as $x$ goes from $-\infty$ to +$\infty$. Thus for the kink solution $\phi^2 \le a^2$. 
In these cases explicit analytic solutions are not possible and we can only 
find implicit kink solutions. From the latter we can obtain 
how a kink profile falls off as $x\rightarrow\pm\infty$.

We will first discuss explicitly the case $n=2$ (i.e. the $\phi^{16}$ field theory) 
and then directly mention results about the kink tail for arbitrary $n$. 
The details about $n=3$ and the generalization to arbitrary $n$ are 
discussed in Appendix E1. For completeness we will also discuss the case of 
$n = 1$ in Appendix E1. It might be noted that
for $n = 1$ we will see that the tails around $\phi = a$ and $\phi = 0$ have
the same asymptotic fall off while for $n \ge 2$ the kink tail around $\phi = a$ has
a slower fall off compared to the tail around $\phi = 0$.

\subsubsection{Case I: $n=2$} 

On using Eq. (\ref{2.3}) the self-dual first order equation is
\be\label{6.8} 
\sqrt{2} \lambda x = \int \frac{d\phi}{\phi^2 (a^2-\phi^2)^3} \,, 
\ee 
The integrand on the right hand side can be evaluated using partial fractions
and it can be shown to be equal to 
\be\label{6.9}
\frac{1}{a^2 (a^2 - \phi^2)^3} + \frac{1}{a^4 (a^2 - \phi^2)^2}
+\frac{1}{a^6 (a^2-\phi^2)} +\frac{1}{a^6 \phi^2}\,.
\ee
This is easily integrated by making use of the identity (\ref{3.4}) leading to
the solution 
\be\label{6.10}
\mu x = \frac{a^3 \phi}{4(a^2-\phi^2)^2} -\frac{a}{\phi} +
\frac{7 a \phi}{8(a^2-\phi^2)} +\frac{15}{16} \ln\frac{a+\phi}{a-\phi}\,.
\ee 
Here $\mu = \sqrt{2} \lambda a^{7}$. Asymptotically, 
\be\label{6.11} 
\lim_{x \rightarrow -\infty} ~\phi(x) = \frac{a}{-\mu x}\,, 
~~~ \lim_{x \rightarrow\infty} ~\phi(x) = a - a\sqrt{\frac{1}{16 \mu x}} \,. 
\ee 
 
Thus the kink tail around $\phi = a$ is entirely determined by the 
first term on the right hand side of Eq. (\ref{6.10}), i.e. by the term
-$\frac{a^3 \phi}{4(a^2-\phi^2)^2}$. On the other hand, the kink tail around 
$\phi = 0$ is entirely determined by the second term on the right hand side of 
Eq. (\ref{6.10}), i.e. by the term $-\frac{a}{\phi}$.

Similar to the $n = 2$ case, one can obtain results for the $n = 3$ case
and then generalize to arbitrary $n$, which are discussed in Appendix E1. As
shown in Appendix E1, for arbitrary $n$, while
the kink tail around $\phi = 0$ always goes like $x^{-1}$, it falls 
off like $x^{-1/n}$ around $\phi = a$.

\subsubsection{Kink Mass}

Using Eq. (\ref{2.5}) one can easily estimate the kink mass for the
entire family of potentials as given by Eq. (\ref{6.2}). We find that 
\be\label{6.16}
M_k = \frac{2^{n+5/2}\lambda a^{2n+5} \Gamma^2(n+2)}{(2n+5) \Gamma(2n+4)}\,,
~~~~n = 1, 2, 3, ...\,.
\ee
For example, while
\be\label{6.17}
M_k(n = 1) = \frac{2^{5/2}}{105} \lambda a^{7}\,,~~
M_k(n = 2) = \frac{2^{5/2}}{315} \lambda a^{9}\,.
\ee
We note that for $a = 1$, the kink mass decreases as $n$ increases. 

\subsection{Models Where Kink Tail Around $\phi = 0$ has a Slower Asymptotic 
Fall off Compared to the Tail Around $\phi = a$}

For simplicity, let us consider the one parameter family of potentials 
\be\label{7.1}
V(\phi) = \lambda^2 \phi^{2n+2}( a^2-\phi^2)^{4}\,,~~~~n = 2,3, ...~\,.
\ee
Similar arguments also hold good in case $(a^2 -\phi^2)^4$ in Eq. (\ref{7.1}) 
is replaced by the potential $(a^2 -\phi^2)^{(2m+2)}$ with $m < n$.

This potential has degenerate minima at $\phi = 0, \pm a$ and 
$V(\phi = 0, \pm a) = 0$ which admits 
a kink from $0$ to $a$ and a mirror kink from $-a$ to $0$ and both kinks
have a power law tail at both the ends.
We  look for a kink solution which oscillates from $0$ to +$a$
as $x$ goes from $-\infty$ to +$\infty$. Thus for the kink solution 
$\phi^2 \le a^2$. 

We consider the two cases of odd $n$ (i.e. potentials of the form $\phi^{4n}$) 
and even $n$ (i.e. potentials of the form $\phi^{4n+2}$) separately. 
In these cases explicit analytic solutions are not possible and we can only 
find implicit kink solutions. From the latter we can obtain 
how a kink profile falls off as $x\rightarrow\pm\infty$.

\subsubsection{$\phi^{4n}$ potentials} 

Consider the class of potentials 
\be\label{7.2} 
V(\phi) = \lambda^2 \phi^{4n}( a^2-\phi^2)^4  \,, ~~~ n = 2, 3, ... ~.
\ee
We consider the case of $n = 2$ and $n=3$ 
and then generalize to arbitrary $n$. We will see that
for $n \ge 2$ the kink tail around $\phi = 0$ has a 
slower fall off compared to the tail around $\phi = a$.

\subsubsection{Case I: $n=2$}   

On using Eq. (\ref{2.3}), the self-dual first order equation is 
\be\label{7.3} 
\frac{d\phi}{dx} = \pm \sqrt{2} \lambda \phi^4 (a^2 - \phi^2)^2\,. 
\ee 
In this case
\be\label{7.4} 
\sqrt{2} \lambda x = \int \frac{d\phi}{\phi^4 (a^2-\phi^2)^2} \,.
\ee 
The integrand on the right hand side can be written as
\be\label{7.5}
\frac{1}{a^4 \phi^4} + \frac{1}{a^4 (a^2 - \phi^2)^2}
+\frac{2}{a^6 \phi^2} + \frac{2}{a^6 (a^2 - \phi^2)}\,.
\ee

This is easily integrated using Eq. (\ref{3.4}) with the solution 
\be\label{7.6} 
\mu x = -\frac{2a^3}{3\phi^3} +\frac{a \phi}{a^2-\phi^2} -\frac{4a}{\phi} 
+\frac{5}{2} \ln\frac{a+\phi}{a-\phi}\,, 
~~~ \mu=2\sqrt{2} \lambda a^7 \,. 
\ee 
Asymptotically, 
\be\label{7.7} 
\lim_{x \rightarrow -\infty} ~\phi(x) = \frac{2^{1/3} a}{(-3\mu x)^{1/3}} \,, 
~~~ \lim_{x \rightarrow\infty} ~\phi(x) = a - \frac{a}{2\mu x}\,. 
\ee 
Thus the kink tail around $\phi = 0$ is entirely determined by the 
first term on the right hand side of Eq. (\ref{7.6}), i.e, the term 
$-\frac{2a^3}{3\phi^3}$. On the other hand, 
the kink tail around $\phi = a$ is entirely determined by the 
second term on the right hand side of Eq. (\ref{7.6}), i.e, the term 
$\frac{a \phi}{a^2 - \phi^2}$.  

Similar to the $n = 2$ case, one can obtain results for the $n = 3$ case
and then generalize to arbitrary $n$ which are discussed in Appendix E2. As
shown in Appendix E2, for arbitrary $n$, while
the kink tail falls off like $x^{-1}$ around $\phi = a$, it falls 
off like $x^{-1/(2n-1)}$ around $\phi = a$ with $n = 2, 3, 4,...$.

\subsubsection{$\phi^{4n+2}$ potentials} 

Consider the class of potentials 
\be\label{8.1} 
V(\phi) = \lambda^2 \phi^{4n+2}( a^2-\phi^2)^4  \,, ~~~ n = 1, 2, 3, ... ~.
\ee
We first discuss the cases of $n =1$ and $n = 2$ 
and then generalize to arbitrary $n$. We will see that
the kink tail around $\phi = 0$ has a 
slower fall off compared to the tail around $\phi = a$.

\subsubsection{Case I: n=1}   

On using Eq. (\ref{2.3}) the self-dual first order equation is
\be\label{8.2} 
\sqrt{2} \lambda x = \int \frac{d\phi}{\phi^3(a^2-\phi^2)^2} \,.  
\ee
The integrand on the right hand side can be evaluated using partial fractions
and it can be shown to be
\be\label{8.3}
\frac{1}{a^4 \phi^3} + \frac{\phi}{a^4 (a^2 - \phi^2)i^2} +\frac{2}{a^6 \phi}
+\frac{2\phi}{a^6 (a^2-\phi^2)}\,.
\ee
This is easily integrated with the solution 
\be\label{8.4} 
\mu x = -\frac{a^2}{\phi^2} +\frac{a^2}{a^2-\phi^2} 
+2 \ln\frac{\phi^2}{a^2-\phi^2} \,, ~~~ \mu=2\sqrt{2} \lambda a^6 \,. 
\ee 
Asymptotically, 
\be\label{8.5} 
\lim_{x \rightarrow -\infty} ~\phi(x) = \frac{a}{\sqrt{-\mu x}} \,, 
~~~ \lim_{x \rightarrow\infty} ~\phi(x) = a - \frac{a}{2 \mu x} \,. 
\ee 
Thus the kink tail around $\phi = 0$ is entirely determined by the 
first term on the right hand side of Eq. (\ref{8.4}), i.e. by the term
-$\frac{a^2}{\phi^2}$. On the other hand, the kink tail around $\phi = a$ is 
entirely determined by the second term on the right hand side of Eq. 
(\ref{8.4}), i.e. by the term $\frac{a^2}{a^2-\phi^2}$.

Similar to the $n = 1$ case, one can obtain results for the $n = 2$ case
and then generalize to arbitrary $n$ which are discussed in Appendix E3. As
shown in Appendix E3, for arbitrary $n$, while
the kink tail falls off like $x^{-1}$ around $\phi = a$, it falls 
off like $x^{-1/2n}$ around $\phi = a$.

On combining the results of the two subsections of Sec. 6.2, it is then clear 
that for the one parameter family of potentials as given by Eq. (\ref{7.1}), 
while the kink tail falls off like $x^{-1}$ around $\phi = a$, it falls off 
like $x^{-1/n}$ around $\phi = 0$ where $n = 2, 3, 4,...$. 

\subsection{Kink Mass}

Using Eq. (\ref{2.5}) one can immediately estimate the kink mass for the
entire family of potentials as given by Eq. (\ref{7.1}). We find that 
\be\label{8.14}
M_k = \frac{2^{7/2}\lambda a^{n+6}}{(n+2)(n+4)(n+6)}\,,
~~~~n = 1, 2, 3, ...\,.
\ee
For example, while
\be\label{8.15}
M_k(n = 1) = \frac{2^{7/2}}{105} \lambda a^{7}\,,~~
M_k(n = 2) = \frac{\sqrt{2}}{24} \lambda a^{8}\,.
\ee
We note that for $a = 1$, the kink mass decreases as $n$ increases. 

\section{Conclusion} 

In this paper we have presented a large variety of potentials which admit
kink solutions with a power law tail at either one end or at both the ends. 
For example, we have presented a one-parameter family of potentials all 
of which admit a kink solution which has a power law tail at both the ends. 
We have considered the linear stability of 
these kink solutions and have shown that in all these cases 
there is no gap between the zero mode and the continuum. This is in contrast 
to the well known kink solutions which have an exponential tail at both the ends
and in which case there is always a gap between the zero mode and the continuum.
As an illustration, in Appendix A we have constructed kink solutions for a 
one parameter family of such potentials, discussed their linear stability 
and shown the existence of the gap between the zero mode and the continuum.
 
We have presented several different classes of models which simultaneously
admit two kink solutions with various different forms of kink tails. For 
example, in Sec. IV we have presented a one-parameter family of potentials 
which admit a kink and a corresponding mirror kink with kink tails of the 
form $eppe$. As mentioned in the introduction, very recently we \cite{cddgkks} 
have calculated the KK and the K-AK force for the one-parameter family 
of potentials discussed in Sec. IV and have shown that while the KK force is
repulsive, the K-AK force is attractive and the magnitude of both these forces
fall off as  $R^{-2(n+1)/n}$ where $R$ is the KK or K-AK separation while
 $n = 1,2,...$. In particular it has been shown 
that
\be\label{h1}
F_{KK} = \bigg [\frac{\Gamma(\frac{n}{2(n+1)}) \Gamma(\frac{1}{2(n+1)})}
{2^{(3n+2)/2(n+1)} n\sqrt{\pi}} \bigg ]^{2(n+1)/n} R^{-2(n+1)/n}\,,
\ee
and 
\be\label{h2}
F_{K-AK} = -\bigg [\frac{\sqrt{\pi} \Gamma(\frac{n}{2(n+1)})}
{2^{n/2(n+1)}\Gamma(\frac{-1}{2(n+1)})} \bigg ]^{2(n+1)/n} R^{-2(n+1)/n}\,,
\ee
so that 
\be\label{h3}
\frac{F_{K-AK}}{F_{K-K}} 
= - \bigg (\sin\left[\frac{\pi}{2(n+1)}\right] \bigg )^{2(n+1)/n}\,. 
\ee       
Notice that as $n$ increases, this ratio goes on decreasing and as 
$n \rightarrow \infty$, the ratio of the K-AK to KK force goes
to zero. In contrast, for the kinks with tails
of the form $eeee$, the magnitude of the K-AK and KK 
forces is always equal.

In Sec. V we have presented a one-parameter family of potentials which gives rise
to a kink and the corresponding mirror kink with the kink tails of the form 
$peep$. Further, in Sec. VI
we have presented two distinct one-parameter family of potentials which admit 
kink and 
mirror kink solutions with the kink tails of the form $pppp$. 
In one family of potentials we have shown that  the 
power law tail around $\phi = 0$ has a slower fall off compared to the tails 
around $\phi = \pm a$ while in the second case it is the opposite.

For completeness, in the Appendices F and G, we have presented two different 
one-parameter family of potentials which admit non-mirror kink solutions with 
kink tails of the form $eeep$ and $pppe$ respectively. 

Finally, we note that our results could provide insight into domain wall properties 
with regard to specific models in high energy physics \cite{Lohe} condensed matter 
physics \cite{Gufan1, Gufan2, Pavlov}  as well as in biology \cite{Boulbitch} and 
cosmology \cite{Greenwood}. 

However, 
there are several open questions which need to be carefully investigated. Some 
of these are as follows, which we hope to address in the near future.  

1. Why is the ratio of the magnitude of the K-AK force to KK force always 
less than one and decreasing monotonically (as $n$ increases) for the kink 
tails of the form $eppe$ as discussed in Sec. IV while it is always 1 for 
the kink tails of the form $eeee$? 

2. What are the KK and the K-AK forces in case the kinks and the
corresponding mirror kinks have kink tails of the form $peep$ as in Sec. V? 
Naively one might think that in this case both the KK and the K-AK forces 
will be exponentially small.
However in view of the power law tails at the two extreme ends, it is not 
obvious to us that this is indeed true. We surise that the KK and the K-AK 
force may in fact turn out to have a power 
law dependence. Another open question is regarding the ratio of the K-AK force
and the KK force in case the kink tails are of the form $peep$. So far as we 
are aware of, until now no one has calculated 
either the KK or the K-AK force in case the kink tails are of the form $peep$. 
We believe that this is a nontrivial question
for which one may require an entirely new formulation. In fact even for the
kink tails of the form $eeep$ as discussed in Appendix F, it is not obvious
to us that either the KK or the K-AK force is exponentially small. This is 
because of the power law tail at one end of one of the kinks. 

3. What about the KK and the K-AK forces when all the four tails are of the 
form $pppp$? There are two possibilities here, both of which have been
discussed  in Sec. VI. In one case
the power law tail around $\phi = \pm a$ has a slower asymptotic 
fall off compared to the asymptotic fall off around $\phi = 0$ while in the 
other case, the power law tail 
around  $\phi = \pm a$ has a faster fall off compared to the 
power law tail around $\phi = 0$. 

In the one parameter family of potentials
$V(\phi) = \lambda^2 \phi^4 (\phi^2 - a^2)^{2n+2}$ with $n = 2, 3, ...$ 
discussed in Sec. VI, while
the kink tail around $\phi = 0$ falls off like $x^{-1}$, while kink tail around
$\phi = a$ falls off like 
$x^{-1/n}$ around $\phi = \pm a$ with $n = 2, 3, ...$. In this case, while
naively one would think that the K-K and the K-AK force will go like $R^{-4}$
where $R$ is the KK or K-AK separation, we believe that because of the kink 
tail around $\phi = \pm a$, the KK and K-AK force may actually fall off 
slower than $R^{-4}$. So far as we are aware of, until now no calculation
exists in the literature for either the KK or the K-AK force in this case. 
We suspect that a conceptually new 
formulation is required to compute KK and K-AK forces in this case.

On the other hand, in the one parameter family of potentials
$V(\phi) = \lambda^2 \phi^{2n+2} (\phi^2 - a^2)^{4}$ with $n = 2, 3, ...$ 
discussed in Sec. VI, the kink tail around $\phi = 0$ falls off like $x^{-1/n}$ 
it falls off like $x^{-1}$ around $\phi = \pm a$ with $n = 2, 3, ...$. In this case, 
we suspect that the KK as well as the K-AK force will indeed go like 
$R^{-2(n+1)/n}$ where $R$ is the KK or K-AK separation where $n = 1,2,3,...$, 
However, we suspect that the strength of the KK and K-AK forces would get 
modified because of the power law tail 
around $\phi = \pm a$. So far as we are aware of, until now no 
calculation exists in the literature about the KK or the K-AK force
in case the kink tails are of the form $pppp$. 
We suspect that even in this case a conceptually new 
formulation is required to compute the KK and the K-AK forces.

Similarly, in the case of the kink tails of the form $pppe$ as discussed in
Appendix G, it is not obvious how the KK or the K-AK force will
vary with $R$ and how the strength of the KK and K-AK force would be modified
by the power law tail at $\phi = -a$. 

4. Can one  rigorously prove our 
assertion that for kinks with a power law tail at either one end or both the
ends, there is no gap between the zero mode and the continuum?


 \section{Acknowledgments} 

We acknowledge fruitful discussions with I. C. Christov, P. G. Kevrekidis and N. S. Manton on higher-order field 
 theories as well as kink interactions in such theories.  A. K. is grateful to INSA (Indian National Science Academy) for the award of INSA Senior 
 Scientist position.  This work was supported in part by the US Department of Energy. 
 
\vskip 0.5truecm 
\noindent{\bf APPENDIX A: Potentials With Exponential Tail At Both Ends}

For completeness, we now discuss a one parameter family of potentials which 
admits a  kink solution as well as a mirror kink solution (and the 
corresponding antikink solutions) 
all of which have an exponential tail at both the ends \cite{manton2}.  
Let us consider the family of potentials 
\be\label{a1} 
V(\phi) = \frac{\lambda^2}{2}\phi^2(\phi^{2n}-a^{2n})^2 \,, ~~~ n=1,2,3 ... \,. 
\ee 
Note that for $n=1$ the existence of such kink solutions is well known 
\cite{Behera, Sanati}.  The self-dual equation is 
\be\label{a2} 
\frac{d\phi}{dx} = \pm \lambda \phi(a^{2n}-\phi^{2n}) \,. 
\ee 
It is easily shown that this equation admits a kink and a mirror kink 
solution, one from $0$ to $a$ and a 
mirror kink from $-a$ to $0$.  The kink solution from $0$ to $a$ is given by 
\be\label{a3} 
\phi(x) = a \left[\frac{(1+\tanh \beta x)}{2}\right]^{1/2n} \,,
\ee 
where $\beta=n\lambda a^{2n}$.  

Hence asymptotically
\be\label{a4} 
\lim_{x \rightarrow-\infty} ~\phi(x) = a e^{\mu x/2n} \,, 
~~~ \lim_{x \rightarrow\infty} ~\phi(x) = a -\frac{a}{4n} e^{-\mu x} \,, ~~~ 
\mu = 2n\lambda a^{2n} \,. 
\ee 
The stability analysis is easily performed.  The corresponding zero mode is 
\be\label{a5} 
\psi_0(x) \propto (1 + \tanh \beta x)^{(\frac{1}{2n} - 1)} \sech^2 \beta x \,, 
\ee 
which is clearly nodeless.  The kink potential is also easily calculated 
\be\label{a6} 
V(x) = V''(\phi_k) = \frac{\lambda^2 a^{4n}}{4} [9(8n^2+6n+1) \sech^2 \beta x 
+ (8n^2-2) \tanh \beta x + (8n^2 + 2) ] \,. \nonumber \\ 
\ee 
Note that  
\be\label{a7} 
V(\infty) = 4 \lambda^2 n^2 a^{2n}\,,  ~~~ V(-\infty) = \lambda^2 a^{2n} \,. 
\ee 
Thus unlike the potentials with a power law tail, in this case with 
exponential tails of the form $eeee$, there is a gap between the zero mode
and the continuum. 
 
\vskip 0.5truecm  
\noindent{\bf APPENDIX B: Kink Solutions for the Potentials Given by 
Eq. (\ref{3.1a})} 

Let us Consider the class of potentials 
\be\label{9.1} 
V(\phi) = \lambda^2 |\phi^2 - a^2|^{2n+1} \,, ~~~ n = 1, 2, 3, ... ~.
\ee
These potentials have degenerate minima at $\phi = \pm a$ and $V(\phi = \pm a)
= 0$. We therefore look for a kink solution which oscillates from $-a$ to +$a$
as $x$ goes from $-\infty$ to +$\infty$. Thus for the kink solution 
$\phi^2 \le a^2$. Notice that while the potential (\ref{9.1}) is continuous its
derivative is not continuous. However since the kink solution oscillates from
$\phi = -a$ to $\phi = +a$ as $x$ goes from $-\infty$ to $\infty$ so that 
$\phi^2 \le a^2$, hence as far as such a kink solution is concerned, our
results would be valid.   

In fact an explicit analytical solution for $n=1$ (i.e. $\phi^6$ field theory) 
has already been obtained previously \cite{Gomes}. Using this explicit solution we
will perform the stability analysis and show that indeed there is no gap between  
the zero mode and the continuum. We also show that even for 
$n=2$ (i.e. $\phi^{10}$ field theory) one can obtain an explicit kink solution 
analytically while for higher values of $n$ one can only get an implicit 
kink solution.  But in all the cases one can determine the asymptotic 
behavior of kink as $x\rightarrow\pm\infty$ and show that there is no
gap between the zero mode and the continuum. 

{\bf Case I:} $n=1$.  
\be\label{9.2} 
V(\phi) = \lambda^2 |\phi^2 - a^2|^3 \,. 
\ee 
Note that the kink solution for this potential has already been obtained 
previously \cite{Gomes}. The self-dual first order equation is 
\be\label{9.3} 
\frac{d\phi}{dx} = \pm \sqrt{2V(\phi)} \,. 
\ee 
This is easily integrated, yielding 
\be\label{9.4} 
\mu x = \frac{\phi}{\sqrt{a^2-\phi^2}} \,, ~~~ \mu = \sqrt{2} \lambda a^2 \,.  
\ee
Thus the kink solution is \cite{Gomes} 
\be\label{9.5} 
\phi(x) = \frac{a \mu x} {\sqrt{1+\mu^2 x^2}} \,.  
\ee 
Asymptotically it behaves as 
\be\label{9.6} 
\lim_{x \rightarrow -\infty} ~\phi(x) = -a + \frac{a}{2\mu^2 x^2} + ... 
\,, ~~~  
\lim_{x \rightarrow \infty} ~\phi(x) = a - \frac{a}{2\mu^2 x^2} + ... \,. 
\ee 

{\bf Case II: $n=2$}.  

In this case the self-dual first order equation is
\be\label{9.7} 
\frac{d\phi}{dx} = \pm \sqrt{2} \lambda |a^2 -\phi^2|^{5/2}\,. 
\ee 
This is easily integrated yielding
\be\label{9.8} 
\mu x = \frac{\phi(3a^2-\phi^2)} {(a^2-\phi^2)^{3/2}} \,,  ~~~ \mu = 3\sqrt{2} \lambda a^4 \,. 
\ee 
On squaring and using $y=\phi^2/a^2$ we get the cubic equation 
\be\label{9.9} 
y^3 -3y^2 + 3\left( \frac{\mu^2 x^2 + 3} {\mu^2 x^2 +4} \right) y - \frac{\mu^2 x^2}{\mu^2x^2+4} = 0 \,. 
\ee 
Its solution is given by 
\be\label{9.10} 
1-y = \frac{a^2-\phi^2}{a^2} =  \left( \frac{\sqrt{\mu^2 x^2 +4} -\mu x} {2(\mu^2 x^2+4)^{3/2}}\right)^{1/3} 
+ \left( \frac{\mu x + \sqrt{\mu^2 x^2 +4}} {2(\mu^2 x^2 +4)^{3/2}} \right)^{1/3} \,. 
\ee 
Thus asymptotically 
\be\label{9.11} 
\lim_{x \rightarrow -\infty} ~\phi(x) = -a + \frac{a}{2(\mu^2 x^2)^{1/3}}\,,
 ~~~  \lim_{x \rightarrow \infty} ~\phi(x) = a - \frac{a}{2(\mu^2 x^2)^{1/3}}\,. 
\ee 

{\bf Case III: General $n$}. 

Consider the potential 
\be\label{9.12} 
V(\phi) = \lambda^2 |\phi^2 - a^2|^{2n+1} \,,  
\ee 
for which, on using the identity (\ref{3.4}) the solution is given by 
\be\label{9.13} 
\mu x = \frac{a^{2n-2}\phi} {(a^2-\phi^2)^{n-1/2}} 
+ \sum_{k=1}^{n-1} \frac{2^k(n-1)...(n-k)} 
{(2n-3)(2n-5)...(2n-2k-1)} \frac{a^{2(n-k-1)}\phi} {(a^2-\phi^2)^{n-k-1/2}} \,, 
\ee 
with 
\be\label{9.14} 
\mu = (2n-1) \sqrt{2} \lambda a^{2n} \,. 
\ee 
Asymptotically the leading contribution as $x\rightarrow\pm\infty$ comes 
from the first term.  We find 
\be\label{9.15} 
\lim_{x \rightarrow -\infty} ~\phi(x) = -a + \frac{a}{2(\mu^2 x^2)^{1/(2n-1)}}
\,, ~~~  \lim_{x \rightarrow \infty} ~\phi(x) = a 
- \frac{a}{2(\mu^2 x^2)^{1/(2n-1)}}\,. 
\ee 
It is clear that this result agrees with that for $n=1,2$.  

Using the exact kink solution for the $\phi^6$ case as given by Eq. (\ref{9.5}) 
 and using Eq. (\ref{2.8}) one can immediately show that in this case the zero mode is 
\be\label{9.16} 
\psi_0 = \frac{d\phi_k}{dx} = \frac{a \mu}{(1+\mu^2 x^2)^{3/2}} \,, 
\ee 
which is indeed nodeless.  Consider the relevant Schr\"odinger-like 
stability equation  
\be\label{9.17} 
-\psi''(x) + V(x) \psi(x) = \omega^2 \psi \,. 
\ee
On using Eq. (\ref{2.7}), the corresponding potential $V(x)$ is given by 
\be\label{9.18} 
V(x) = \frac{d^2V}{d\phi^2}\Big|_{\phi = \phi_k} = \frac{-3\mu^2(1 
-4 \mu^2 x^2)}{(1+\mu^2 x^2)^2} \,. 
\ee
This potential vanishes at $x = \pm \infty$, has a minimum at $x = 0$ with 
$V_{min} = -3\mu^2$ and has maxima at $\mu^2 x^2 = 1$ with $V = 9/4$.  
Note that $V$ vanishes at $x = \pm \infty$, and hence 
the continuum begins at the same energy as that of the zero 
mode. This is quite different from the case of the kinks with exponential tail 
at both the ends in which there is always a gap between the zero mode and 
the continuum. 
Even for the kink solution of the $\phi^{10}$ field theory  as given by 
Eq. (\ref{9.10}), we have 
checked that the corresponding $V(x)$ of the stability Eq. (\ref{9.17}) 
vanishes at $x = \pm\infty$ 
and hence the continuum begins at the same energy as that of the zero mode, 
i.e. there is no gap between the zero mode and the continuum. 
These two examples thus support the fact that unlike 
the kinks with exponential tails at both the ends, for the kink solution 
with a power law tail at either one or both the ends, there is no gap between 
the zero mode and the continuum.

\vskip 0.5truecm  
\noindent{\bf APPENDIX C: Details About Kink Solutions of Eq. (\ref{4.17}) 
of Sec. IV} 

We now discuss the kink solutions for a class of potentials given by 
Eq. (\ref{4.17}). We have already discussed the $n = 1$ case in Sec. IV and
we now discuss the $n =2$ case and then generalize to arbitrary $n$.
 
{\bf Case II: $n=2$} 

On using Eq. (\ref{2.3}) the self-dual first order equation is
\be\label{4.22} 
\sqrt{2} \lambda x = \int \frac{d\phi}{\phi^5(a^2-\phi^2)} \,, 
\ee 
The integrand on the right hand side can be evaluated using partial fractions
and it can be shown to be
\be\label{4.23}
\frac{1}{a^2 \phi^5} + \frac{1}{a^4 \phi^3} +\frac{1}{a^6 \phi} 
+ \frac{\phi}{a^6 (a^2 - \phi^2)}\,.
\ee
This is easily integrated with the solution 
\be\label{4.24}
\mu x = -\frac{a^4}{2\phi^4}  - \frac{a^2}{\phi^2} 
+ \ln\frac{\phi^2}{a^2-\phi^2} \,, ~~~ \mu=2\sqrt{2} \lambda a^6 \,. 
\ee 
Asymptotically, 
\be\label{4.25} 
\lim_{x \rightarrow -\infty} ~\phi(x) = \frac{a}{(-2\mu x)^{1/4}} \,, 
~~~ \lim_{x \rightarrow +\infty} ~\phi(x) 
= a - \frac{a}{2} e^{-\mu x-1-1/2} \,. 
\ee 
Thus the kink tail around $\phi = 0$ is again entirely determined by the 
first term on the right hand side of Eq. (\ref{4.24}), i.e. by the term
-$\frac{a^4}{2\phi^4}$. Further, on comparing the kink tail around $\phi 
= a$ (i.e. as $x \rightarrow +\infty$) in the case of $n = 1$ and $n = 2$ 
as given by Eqs. (\ref{4.21}) and (\ref{4.25}) respectively, we find that in 
both the cases the kink tail has exponential fall off
\be\label{4.26}
\lim_{x \rightarrow \infty}~ \phi(x) = a - \frac{a}{2} e^{-\mu x - K}\,,
\ee
where the constant $K$ equals $1$ in the $n = 1$ case while it is $1+1/2$ in
the $n = 2$ case and the rest of the behaviour is the same in both the cases.

{\bf Case III: General $n$}

The generalization to arbitrary $n$ is now straightforward.
On using Eq. (\ref{2.3}), the self-dual first order equation is 
\be\label{4.27}
\sqrt{2} \lambda x = \int \frac{d\phi}{\phi^{2n+1}(a^2-\phi^2)} \,. 
\ee 
The integrand on the right hand side can be evaluated using partial fractions
and it can be shown to be
\be\label{4.28}
\sum_{k=1}^{n+1} \frac{1}{a^{2k} \phi^{2n+3-2k}}  
+\frac{\phi}{a^{2n+2} (a^2 - \phi^2)}\,.
\ee
This is easily integrated with the solution 
\be\label{4.29} 
\mu x = - \frac{a^{2n}}{n\phi^{2n}} 
-\sum_{k=2}^{n} \frac{a^{2n+2-2k}}{(n+1-k)\phi^{2n+2-2k}} 
+ \ln\frac{\phi^2}{a^2-\phi^2} \,. 
\ee
Asymptotically, 
\be\label{4.30} 
\lim_{x \rightarrow -\infty} ~\phi(x) = \frac{a}{(-n\mu x)^{1/2n}} \,, 
~~~ \lim_{x \rightarrow +\infty} ~\phi(x) = a - \frac{a}{2} e^{-\mu x-D_n} \,,  
\ee 
where $D_n = 1 +1/2 + 1/3 + ... + 1/n$. 

As expected, for $n=1,2$ these results for the kink tail agree with those given by
Eqs. (\ref{4.21}) and (\ref{4.25}), respectively. 

\vskip 0.5truecm  
\noindent{\bf APPENDIX D: Details About Kink Solutions of Eq. (\ref{5.1}) in  
Sec. V} 

We now discuss the kink solution for the class of potentials given by 
Eq. (\ref{5.1}). We have already discussed the $n = 1$ case in Sec. V and
we now discuss the $n =2$ case and then generalize to arbitrary $n$.
 
{\bf Case II: $n=2$} 

On using Eq. (\ref{2.3}) the self-dual first order equation is
\be\label{5.7} 
\sqrt{2} \lambda x = \int \frac{d\phi}{\phi(a^2-\phi^2)^3} \,.  
\ee 
The integrand on the right hand side can be evaluated using partial fractions
and it can be shown to be qual to 
\be\label{5.8}
 \frac{\phi}{a^2 (a^2 - \phi^2)^3} 
+\frac{\phi}{a^4 (a^2-\phi^2)^2} +\frac{\phi}{a^6 (a^2-\phi^2)}
+\frac{1}{a^6 \phi}\,. 
\ee
This is easily integrated with the 
solution 
\be\label{5.9}
\mu x = \frac{a^4}{2(a^2-\phi^2)^2}+\frac{a^2}{(a^2-\phi^2)}
\mu x = \ln\frac{\phi^2}{(a^2-\phi^2)}\,,~~~ \mu =2 \sqrt{2} \lambda a^6 \,.  
\ee 
Asymptotically, 
\be\label{5.10} 
\lim_{x \rightarrow -\infty} ~\phi(x) = a e^{\mu x /2-2-1/2} \,, ~~~
\lim_{x \rightarrow \infty} ~\phi(x) = a-\frac{a}{(8\mu x)^{1/2}} \,.  
\ee 
Thus the kink tail around $\phi = a$ is again entirely determined by the 
first term on the right hand side of Eq. (\ref{5.9}), i.e. by the term
-$\frac{a^4}{2(a^2-\phi^2)^2}$. Further, on comparing the kink tail around 
$\phi = 0$ (i.e. as $x \rightarrow -\infty$) in the case of $n = 1$ and $n = 2$ 
as given by Eqs. (\ref{5.6}) and (\ref{5.10}) respectively, we find that 
in both the cases the kink tail has exponential fall
\be\label{5.11}
\lim_{x \rightarrow -\infty} ~ \phi(x) = a e^{\mu x /2 - K}\,,
\ee
where the constant $K$ equals $1$ in the $n = 1$ case while it is 
$1+\frac{1}{2}$ in the $n = 2$ case and the rest of the behaviour is the 
same in both the cases.

{\bf Case III: General $n$} 

The generalization to arbitrary $n$ is now straightforward.
On using Eq. (\ref{2.3}), the self-dual first order equation is 
\be\label{5.12}
\sqrt{2} \lambda x = \int \frac{d\phi}{\phi (a^2-\phi^2)^{n+1}} \,. 
\ee 
The integrand on the right hand side can be evaluated using partial fractions
and it can be shown to be equal to 
\be\label{5.13}
\sum_{k=1}^{n+1} \frac{\phi}{a^{2k} (a^2-\phi^2)^{n+2-k}} 
+\frac{1}{a^{2n+2} \phi}\,.
\ee
This is easily integrated with the solution 
\be\label{5.14} 
\mu x = \frac{a^{2n}}{n(a^2 -\phi^2)^{n}}
+\sum_{k=2}^{n}\frac{a^{2n+2-2k}}{(n+1-k) (a^2-\phi^2)^{n+1-k}} 
+ \ln\frac{\phi^2}{(a^2-\phi^2)}\,. 
\ee
Asymptotically, 
\be\label{5.15} 
\lim_{x \rightarrow -\infty} ~\phi(x) = a e^{\mu x /2 - D_n} \,, 
~~~ \lim_{x \rightarrow +\infty} ~\phi(x) = a 
- \frac{a}{(n 2^{n}\mu x)^{1/n}} \,,  
\ee 
where $D_n = 1 +1/2 + 1/3 + ... + 1/n$. 

As expected, for $n=1,2$ these results for the kink tail agree with those given  
by Eqs. (\ref{5.6}) and (\ref{5.10}), respectively.

\vskip 0.5truecm  
\noindent{\bf APPENDIX E1: Details About Kink Solutions of Eq. (\ref{6.2}) in
Sec. VI} 

We now discuss the kink solutions for the class of potentials given by 
Eq. (\ref{6.2}). We have already discussed the $n = 2$ case in Sec. VI and
we now discuss the $n =3$ case and then generalize to arbitrary $n$. For
completeness we first discuss the $n=1$ case in which the kink tail
around $\phi = a$ as well as around $\phi = 0$ both fall off like $x^{-1}$.

{\bf Case II: $n=1$}   

On using Eq. (\ref{2.3}) in Eq. (\ref{6.2}), for $n = 1$ the self-dual first 
order equation is 
\be\label{6.3} 
\frac{d\phi}{dx} = \pm \sqrt{2} \lambda \phi^2 (a^2 - \phi^2)^2\,. 
\ee 
In this case
\be\label{6.4} 
\sqrt{2} \lambda x = \int \frac{d\phi}{\phi^2 (a^2-\phi^2)^2} \,.
\ee 
The integrand on the right hand side can be written as
\be\label{6.5}
\frac{1}{a^2 (a^2 - \phi^2)^2} + \frac{1}{a^4 (a^2 - \phi^2)}
+\frac{1}{a^4 \phi^2}\,.
\ee
This is easily integrated using the identity (\ref{3.4}) with the solution 
\be\label{6.6} 
\mu x = \frac{a\phi}{2(a^2 - \phi^2)} -\frac{a}{\phi} 
+\frac{3}{4}\ln\frac{a+\phi}{a-\phi}\,, 
~~~ \mu= \sqrt{2} \lambda a^5\,. 
\ee 
Asymptotically, 
\be\label{6.7} 
\lim_{x \rightarrow -\infty} ~\phi(x) = \frac{a}{-\mu x}\,, 
~~~ \lim_{x \rightarrow\infty} ~\phi(x) = a - \frac{a}{4\mu x} \,. 
\ee 
Thus while the kink tail around $\phi = a$ is entirely determined by the 
first term on the right hand side of Eq. (\ref{6.6}), i.e, by the term 
$\frac{a\phi}{2(a^2 -\phi^2)}$, the kink tail around $\phi = 0$ is entirely
decided by the second term on the right hand side of Eq. (\ref{6.6}), i.e.
by the term -$\frac{a}{\phi}$. Observe that in this case the kink tail 
falls off like $x^{-1}$ around both $\phi = a$ and $\phi =0$. 

{\bf Case III: $n = 3$}

On using Eq. (\ref{2.3}) in Eq. (\ref{6.2}), for $n = 3$ the self-dual first 
order equation is 
\be\label{6.3a} 
\frac{d\phi}{dx} = \pm \sqrt{2} \lambda \phi^2 (a^2 - \phi^2)^4\,. 
\ee 
In this case
\be\label{6.4a} 
\sqrt{2} \lambda x = \int \frac{d\phi}{\phi^2 (a^2-\phi^2)^4} \,.
\ee 
The integrand on the right hand side can be written as
\be\label{6.5a}
\frac{1}{a^2 (a^2 - \phi^2)^4} + \frac{1}{a^4 (a^2 - \phi^2)^3}
+\frac{1}{a^{6} (a^2 - \phi^2)}+\frac{1}{a^6 \phi^2}\,.
\ee
This is easily integrated using the identity (\ref{3.4}) with the solution 
\be\label{6.6a} 
\mu x = \frac{a^{5}\phi}{6(a^2 - \phi^2)^3} -\frac{a}{\phi} 
+O(\frac{\phi}{(a^2-\phi^2)^{p}})+ K\ln\frac{a+\phi}{a-\phi}\,, 
~~~ \mu= \sqrt{2} \lambda a^9\,.~~p =1,2\,. 
\ee 
It may be noted that we have only 
explicitly specified those terms which are relevant for
knowing the kink tail around $\phi = a$ as well as around $\phi = 0$.
Asymptotically, 
\be\label{6.7a} 
\lim_{x \rightarrow -\infty} ~\phi(x) = \frac{a}{-\mu x}\,, 
~~~ \lim_{x \rightarrow\infty} ~\phi(x) = a - \frac{a}{(48\mu x)^{1/3}} \,. 
\ee 
Thus while the kink tail around $\phi = a$ is entirely determined by the 
first term on the right hand side of Eq. (\ref{6.6a}), i.e, by the term 
$\frac{a^5\phi}{2(a^2 -\phi^2)^3}$, the kink tail around $\phi = 0$ is entirely
decided by the second term on the right hand side of Eq. (\ref{6.6a}), i.e.
by the term -$\frac{a}{\phi}$. Observe that in this case the kink tail 
falls off like $x^{-1}$ around  $\phi = 0$ while it falls off like
$x^{-1/3}$ around  $\phi = a$.

{\bf Case IV: General $n$} 

The generalization to arbitrary $n$ is now straightforward.
On using Eq. (\ref{2.3}), the self-dual first order equation is 
\be\label{6.12}
\sqrt{2} \lambda x = \int \frac{d\phi}{\phi^2 (a^2-\phi^2)^{n+1}} \,. 
\ee 
The integrand on the right hand side can be evaluated using partial fractions
and it can be shown to be
\be\label{6.13}
\sum_{k=1}^{n+1}\frac{1}{a^{2k} (a^2-\phi^2)^{n+2-k}} 
+\frac{1}{a^{2n+2} \phi^2}\,.
\ee
This is easily integrated using the identity (\ref{3.4}) with the solution 
\be\label{6.14} 
\mu x = \frac{a^{2n-1} \phi}{2n(a^2 -\phi^2)^{n}} -\frac{a}{\phi}
+O(\frac{\phi}{(a^2-\phi^2)^{p}}) + K \ln\frac{a+\phi}{a-\phi}\,,
\ee
where $\mu = \sqrt{2} \lambda a^{2n+3}$ while $p = 1, 2,...,(n-1)$. 
It may be noted that we have only 
explicitly specified those terms which are relevant for
knowing the kink tail around $\phi = a$ as well as around $\phi = 0$.

Asymptotically, 
\be\label{6.15} 
\lim_{x \rightarrow-\infty} ~\phi(x) = \frac{a}{-\mu x}\,, 
~~~ \lim_{x \rightarrow\infty} ~\phi(x) = a 
- \frac{a}{(n 2^{n+1}\mu x)^{1/n}} \,.
\ee
As expected, for $n=2, 3$ these results for the kink tail agree with those given 
by Eqs. (\ref{6.10}) and (\ref{6.7a}),  respectively. 

\vskip 0.5truecm  
{\bf APPENDIX E2: Details About Kink Solutions of Eq. (\ref{7.2}) in Sec. VI} 

We now discuss the kink solution for the class of potentials given by 
Eq. (\ref{7.2}). We have already discussed the $n = 2$ case in Sec. VI and
we now discuss the $n =3$ case and then generalize to arbitrary $n$. 

{\bf Case II: $n = 3$}

On using Eq. (\ref{2.3}), the self-dual first order equation is 
\be\label{7.8} 
\sqrt{2} \lambda x = \int \frac{d\phi}{\phi^6 (a^2-\phi^2)^2} \,. 
\ee 
The integrand on the right hand side can be evaluated using partial fractions
and it can be shown to be
\be\label{7.9}
\frac{1}{a^4 \phi^6} + \frac{1}{a^6 (a^2 - \phi^2)^2} +\frac{2}{a^6 \phi^4}
+\frac{3}{a^8 \phi^2} +\frac{3}{a^{8} (a^2-\phi^2)}\,.
\ee
This is easily integrated with the solution 
\be\label{7.10}
\mu x = - \frac{2a^5}{5\phi^5} +\frac{a \phi}{a^2-\phi^2} -\frac{4a^3}{3\phi^3} 
-\frac{6a}{\phi}+\frac{7}{2} \ln\frac{a+\phi}{a-\phi} \,, 
~~~ \mu=2\sqrt{2} \lambda a^9 \,. 
\ee 
Asymptotically, 
\be\label{7.11} 
\lim_{x \rightarrow-\infty} ~\phi(x) = \frac{2 a}{(-5\mu x)^{1/5}} \,,
 ~~~ \lim_{x \rightarrow\infty} ~\phi(x) = a - \frac{a}{2 \mu x}\,. 
\ee 
Thus the kink tail around $\phi = 0$ is again entirely determined by the 
first term on the right hand side of Eq. (\ref{7.10}), i.e. by the term
-$\frac{2a^5}{\phi^5}$. 
On the other hand, the kink tail around $\phi = a$ is entirely determined by 
the second term on the right hand side of Eq. (\ref{7.10}), i.e, by the term 
$\frac{a \phi}{a^2 - \phi^2}$. It may be noted that the behaviour of the kink 
tail around $\phi = a$ (i.e. when $x \rightarrow \infty$) is the same for 
$n = 1, 2, 3$.  

It is then clear that even for arbitrary $n$, the behaviour of the kink tail
around $\phi = a$ will be the same as for $n = 1, 2, 3$. On the other hand
the behaviour of the kink tail around $\phi = 0$ (i.e. when 
$x \rightarrow 0$) will simply depend on the most singular
term in $\phi$ as $\phi \rightarrow 0$. 

{\bf Case III: General $n$} 

The generalization to arbitrary $n$ is now straightforward.
On using Eq. (\ref{2.3}), the self-dual first order equation is 
\be\label{7.12}
\sqrt{2} \lambda x = \int \frac{d\phi}{\phi^{2n}(a^2-\phi^2)^2} \,. 
\ee 
The integrand on the right hand side can be evaluated using partial fractions
and it can be shown to be
\be\label{7.13}
\sum_{k=1}^{n} \frac{n-k+1}{a^{2(n-k+2)} \phi^{2k}} 
+\frac{1}{a^{2n} (a^2 - \phi^2)^2} +\frac{n}{a^{2n+2} (a^2-\phi^2)}\,.
\ee
This is easily integrated with the solution 
\be\label{7.14} 
\mu x = - \frac{2a^{2n-1}}{(2n-1)\phi^{2n-1}} +\frac{a\phi}{a^2-\phi^2}
-\sum_{k =1}^{n-1} \frac{2 (n-k+1) a^{2k-1}}{(2k-1)\phi^{2k-1}} 
+\frac{(2n+1)}{2} \ln\frac{a+\phi}{a-\phi} \,,
\ee
where
\be\label{7.15}
\mu=2\sqrt{2} \lambda a^{2n+3} \,. 
\ee 
Asymptotically, 
\be \label{7.16}
\lim_{x \rightarrow-\infty} ~\phi(x) = \frac{2^{1/{(2n-1)}} a}
{[-(2n-1)\mu x]^{1/(2n-1)}} \,, 
~~~ \lim_{x \rightarrow\infty} ~\phi(x) = a - \frac{a}{2\mu x}\,.  
\ee 

As expected, for $n=1,2,3$ these results for the kink tail agree with those given by
Eqs. (\ref{6.7}), (\ref{7.7}) and (\ref{7.11}), respectively. Thus the 
kink tail around $\phi = a$ goes like $x^{-1}$ while it goes like 
$x^{-1/(2n-1)}$ around $\phi = 0$.

\vskip 0.5truecm  
{\bf Appendix E3: Details About Kink Solutions of Eq. (\ref{8.1}) in Sec. VI} 

We now discuss the kink solutions for a class of potentials given by 
Eq. (\ref{8.1}). We have already discussed the $n = 1$ case in Sec. VI and
we now discuss the $n =2$ case and then generalize to arbitrary $n$. 

{\bf Case II: $n = 2$}

On using Eq. (\ref{2.3}) the self-dual first order equation is
\be\label{8.6} 
\sqrt{2} \lambda x = \int \frac{d\phi}{\phi^5(a^2-\phi^2)^2} \,.  
\ee 
The integrand on the right hand side can be evaluated using partial fractions
and it can be shown to be
\be\label{8.7}
\frac{1}{a^4 \phi^5} + \frac{\phi}{a^6 (a^2-\phi^2)^2} +\frac{2}{a^6 \phi^3} 
+\frac{3}{a^8 \phi} + \frac{3\phi}{a^8 (a^2 - \phi^2)}\,.
\ee
This is easily integrated with the solution 
\be\label{8.8}
\mu x = -\frac{a^4}{2\phi^4} +\frac{a^2}{a^2-\phi^2}  - \frac{2 a^2}{\phi^2} 
+3 \ln\frac{\phi^2}{a^2-\phi^2} \,, ~~~ \mu=2\sqrt{2} \lambda a^8 \,. 
\ee 
Asymptotically, 
\be\label{8.9} 
\lim_{x \rightarrow-\infty} ~\phi(x) = \frac{a}{(-2\mu x)^{1/4}} \,, 
~~~ \lim_{x \rightarrow\infty} ~\phi(x) = a - \frac{a}{2 \mu x} \,. 
\ee 
Thus the kink tail around $\phi = 0$ is again entirely determined by the 
first term on the right hand side of Eq. (\ref{8.8}), i.e. by the term
-$\frac{a^4}{2\phi^4}$. On the other hand, the kink tail around $\phi 
= a$ is again determined by the second term on the right hand side of 
Eq. (\ref{8.8}), i.e. by the term $\frac{a^2}{a^2-\phi^2}$. It is
worth pointing out that the kink tail around $\phi = a$ is the {\it same}
in the case of $n = 1$ and $n = 2$, and we shall see below that it is the 
same for any $n$. 

{\bf Case III: General $n$} 

The generalization to arbitrary $n$ is now straightforward.
On using Eq. (\ref{2.3}), the self-dual first order equation is 
\be\label{8.10}
\sqrt{2} \lambda x = \int \frac{d\phi}{\phi^{2n+1}(a^2-\phi^2)^2} \,. 
\ee 
The integrand on the right hand side can be evaluated using partial fractions
and it can be shown to be
\be\label{8.11}
+\sum_{k=1}^{n+1} \frac{n-k+2}{a^{2(n-k+3)} \phi^{2k-1}}  
+\frac{\phi}{a^{2n+2} (a^2 - \phi^2)^2} 
+\frac{(n+1)\phi}{a^{2n+4}(a^2-\phi^2)}\,.
\ee
This is easily integrated with the solution 
\be\label{8.12} 
\mu x = - \frac{a^{2n}}{n\phi^{2n}}+\frac{a^2}{a^2-\phi^2}  
-\sum_{k=2}^{n} \frac{(n+2-k) a^{2(k-1)}}{(k-1)\phi^{2(k-1)}} 
+ (n+1)\ln\frac{\phi^2}{a^2-\phi^2} \,,
\ee
where $\mu = 2\sqrt{2} \lambda a^{2n+4}$. Asymptotically, 
\be\label{8.13} 
\lim_{x \rightarrow-\infty} ~\phi(x) = \frac{a}{(-n\mu x)^{1/2n}} \,, 
~~~ \lim_{x \rightarrow\infty} ~\phi(x) = a - \frac{a}{2 \mu x} \,.  
\ee 

As expected, for $n=1,2$ these results for the kink tail agree with those 
given by Eqs. (\ref{8.5}) and (\ref{8.9}), respectively. 

\vskip 0.5truecm  
{\bf APPENDIX F: Potentials admitting kink solutions with tails of 
the form $e~e~e~p$}

We now briefly discuss a family of potentials of the form
\be\label{f.1}
V(\phi) = \lambda^2 (a^2 - \phi^2)^2 (b^2-\phi^2)^{2n+2}  \,, ~~b > a\,,
~~~ n = 1, 2, 3, ... ~\,.
\ee
These potentials have degenerate minima at $\phi = \pm a, \pm b$ and 
$V(\phi = \pm a, \pm b) = 0$ and admit 
a kink from $-a$ to $a$ and a kink from $a$ to $b$ as well as a mirror kink from
$-b$ to $-a$. While the kink from $-a$ to $a$ has an exponential tail around both
$\phi = -a$ as well as $\phi = a$, the kink from $a$ to $b$
has an exponential tail around $\phi = a$ and a power law tail around $\phi = b$.
In these cases too explicit analytic solutions are 
not possible and we can only 
find implicit kink solutions. From the latter we can obtain 
how a kink profile falls off as $x\rightarrow\pm\infty$.  

{\bf $-a$ to $a$ kink}

On using Eq. (\ref{2.3}), the self-dual first order equation for the potential
(\ref{f.1}) is 
\be\label{f.2}
\sqrt{2} \lambda x = \int \frac{d\phi}{(a^2 -\phi^2) (b^2-\phi^2)^{n+1}} \,. 
\ee 
The integrand on the right hand side can be evaluated using partial fractions
and it can be shown to be
\be\label{f.3}
+\sum_{k=1}^{n+1} \frac{1}{(b^2-a^2)^{(n-k+2)}  (b^2-\phi^2)^{k}} 
+\frac{1}{(b^2 -a^2)^{n+1} (a^2 - \phi^2)}\,.
\ee
This is easily integrated with the solution 
\be\label{f.4} 
\mu x = \ln (\frac{a+\phi}{a-\phi}) - \frac{a \phi (b^2-a^2)^{n}}
{n b^2 (b^2 -\phi^2)^{n}} + Lower~Order~Terms\,,
\ee
where $\mu = 2 \sqrt{2} a \lambda (b^2 -a^2)^{n+1}$. 
Note that in Eq. (\ref{f.4}) we have only specified those terms which 
contribute to the dominant asymptotic behaviour as $x \rightarrow \pm \infty$.  
Asymptotically we find that
\be\label{f.5} 
\lim_{x \rightarrow -\infty} ~\phi(x) = -a + 2a f(a,b) e^{\mu x} \,, 
~~~ \lim_{x \rightarrow +\infty} ~\phi(x) = a 
- 2a f(a, b) e^{-\mu x} \,.  
\ee 
Thus for the kink solution from $-a$ to $a$, the kink tail around both 
$\phi = -a$ and $\phi = a$ has an exponential tail. 

{\bf $a$ to $b$ kink}

On using Eq. (\ref{2.3}), the self-dual first order equation is now given by 
\be\label{f.6}
\sqrt{2} \lambda x = \int \frac{d\phi}{(\phi^2- a^2) (b^2-\phi^2)^{n+1}} \,. 
\ee 
The integrand on the right hand side can be evaluated using partial fractions
and it can be shown to be
\be\label{f.7}
+\sum_{k=1}^{n+1} \frac{1}{(b^2-a^2)^{(n-k+2)}  (b^2-\phi^2)^{k}} 
+\frac{1}{(b^2 -a^2)^{n+1} (\phi^2- a^2)}\,.
\ee
This is easily integrated with the solution 
\be\label{f.8} 
\mu x = \ln (\frac{\phi-a}{\phi+a}) + \frac{a \phi (b^2-a^2)^{n}}
{n b^2 (b^2 -\phi^2)^{n}} + Lower~Order~Terms\,,
\ee
where $\mu = 2 \sqrt{2} a \lambda (b^2 -a^2)^{n+1}$. 
Note that in Eq. (\ref{f.8}) we have only specified those terms which 
contribute to the dominant asymptotic behaviour as $x \rightarrow \pm \infty$.  
Asymptotically, 
\bea\label{f.9} 
&&\lim_{x \rightarrow -\infty} ~\phi(x) = a - 2a f(a,b) e^{\mu x}\,, 
\nonumber \\ 
&&\lim_{x \rightarrow +\infty} ~\phi(x) = b 
- b \bigg [\frac{a(b^2-a^2)^{n}}{n 2^{n} b^{2n+1} \mu x} \bigg ]^{1/n} \,. 
\eea 
Thus for the kink solution from $a$ to $b$, while the kink tail around 
$\phi = a$ has an exponential tail, the kink tail around $\phi = b$ goes like
$x^{-1/n}$ as $x \rightarrow  \infty$

\vskip 0.5truecm  
{\bf APPENDIX G: Potentials admitting kink solutions with tails of 
the form $p~p~p~e$}

We now briefly discuss a family of potentials of the form
\be\label{g.1}
V(\phi) = \lambda^2 (a^2 - \phi^2)^{n+1} (b^2-\phi^2)^{2}  \,, ~~b > a\,,
~~~ n = 1, 2, 3, ... ~\,.
\ee
These potentials have degenerate minima at $\phi = \pm a, \pm b$ and 
$V(\phi = \pm a, \pm b) = 0$ and admit 
a kink from $-a$ to $a$ and a kink from $a$ to $b$ as well as a mirror kink from
$-b$ to $-a$. While the kink from -$a$ to $a$ has a power law tail around both
$\phi = -a$ as well as $\phi = a$, the kink from $a$ to $b$
has a power law tail around $\phi = a$ and an exponential tail around $\phi = b$.
In these cases too explicit analytic solutions are 
not possible and we can only 
find implicit kink solutions. From the latter we can obtain 
how a kink profile falls off as $x\rightarrow\pm\infty$.  

{\bf $-a$ to $a$ kink}

On using Eq. (\ref{2.3}), the self-dual first order equation for the potential
(\ref{g.1}) is 
\be\label{g.2}
\sqrt{2} \lambda x = \int \frac{d\phi}{(b^2 -\phi^2) (a^2-\phi^2)^{n+1}} \,. 
\ee 
The integrand on the right hand side can be evauated using partial fractions
and it can be shown to be
\be\label{g.3}
+\sum_{k=1}^{n+1} \frac{(-1)^{n-k+2}}{(b^2-a^2)^{(n-k+2)}  (a^2-\phi^2)^{k}} 
+\frac{(-1)^{n+1}}{(b^2 -a^2)^{n+1} (b^2 - \phi^2)}\,.
\ee
This is easily integrated with the solution 
\be\label{g.4} 
\mu x = \frac{b \phi (b^2-a^2)^{n}}{n a^2 (a^2 -\phi^2)^{n}} 
+ (-1)^{n+1} \ln (\frac{b+\phi}{b-\phi}) + Lower~Order~Terms\,,
\ee
where $\mu = 2 \sqrt{2} b \lambda (b^2 -a^2)^{n+1}$. 
Note that in Eq. (\ref{g.4}) we have only specified those terms which 
contribute to the dominant asymptotic behaviour as $x \rightarrow \pm \infty$.  
Asymptotically, 
\bea\label{g.5} 
&&\lim_{x \rightarrow -\infty} ~\phi(x) = -a 
+ a \bigg [\frac{-b(b^2-a^2)^n}{n 2^n a^{2n+1} \mu x} \bigg ]^{1/n}\,, 
\nonumber \\
&&\lim_{x \rightarrow +\infty} ~\phi(x) = a 
- a \bigg [\frac{b(b^2-a^2)^n}{n 2^n a^{2n+1} \mu x} \bigg ]^{1/n}\,.
\eea
Thus for the kink solution from $-a$ to $a$, the kink tail around both 
$\phi = -a$ and $\phi = a$ has a power law tail..

{\bf $a$ to $b$ kink}

On using Eq. (\ref{2.3}), the self-dual first order equation is now given by 
\be\label{g.6}
\sqrt{2} \lambda x = \int \frac{d\phi}{(b^2 -\phi^2) (\phi^2 -a^2)^{n+1}} \,. 
\ee 
The integrand on the right hand side can be evaluated using partial fractions
and it can be shown to be
\be\label{g.7}
+\sum_{k=1}^{n+1} \frac{1}{(b^2-a^2)^{(n-k+2)}  (\phi^2 -a^2)^{k}} 
+\frac{1}{(b^2 -a^2)^{n+1} (b^2 - \phi^2)}\,.
\ee
This is easily integrated with the solution 
\be\label{g.8} 
\mu x = \ln (\frac{b+\phi}{b-\phi}) - \frac{b \phi (b^2-a^2)^{n}}
{n a^2 (\phi^2 -a^2)^{n}} + Lower~Order~Terms\,,
\ee
where $\mu = 2 \sqrt{2} a \lambda (b^2 -a^2)^{n+1}$. 
Note that in Eq. (\ref{g.8}) we have only specified those terms which 
contribute to the dominant asymptotic behaviour as $x \rightarrow \pm \infty$.  
Asymptotically, 
\bea\label{g.9} 
&&\lim_{x \rightarrow -\infty} ~\phi(x) = a 
+ a \bigg [\frac{-b(b^2-a^2)^n}{n 2^n a^{2n+1} \mu x} \bigg ]^{1/n}\,, 
\nonumber \\ 
&&\lim_{x \rightarrow +\infty} ~\phi(x) = b -2b h(a,b) e^{-\mu x}\,.
\eea 
Thus for the kink solution from $a$ to $b$, while the kink tail around 
$\phi = b$ is exponential, the kink tail around $\phi = a$ goes like
$(-x)^{-1/n}$ as $x \rightarrow  -\infty$.

\end{document}